\documentclass[11pt]{JHEP3}

\JHEPspecialurl{http://jhep.sissa.it/JOURNAL/JHEP3.tar.gz}

\usepackage{epsfig,multicol}
\usepackage{graphicx}
\usepackage{amsmath,amssymb}
\usepackage{mathrsfs}

\newcommand\fverb{\setbox\fverbbox=\hbox\bgroup\verb}
\newcommand\fverbdo{\egroup\medskip\noindent%
			\fbox{\unhbox\fverbbox}\ }
\newcommand\fverbit{\egroup\item[\fbox{\unhbox\fverbbox}]}
\newbox\fverbbox

\newcommand{\nn}{\nonumber}
\def\dfrac#1#2{\displaystyle\frac{#1}{#2}}

\newcommand{\pslash}{p\kern-1ex /}
\newcommand{\qslash}{q\kern-1ex /}
\newcommand{\lslash}{l\kern-1ex /}
\newcommand{\sslash}{s\kern-1ex /}
\newcommand{\kaslash}{k_a\kern-2ex /}
\newcommand{\kbslash}{k_b\kern-2ex /}
\newcommand{\Dslash}{\mathcal{D}\kern-1.5ex /}

\newcommand{\beqa}{\begin{eqnarray}}
\newcommand{\eeqa}{\end{eqnarray}}

\newcommand{\ba}{\begin{eqnarray}}
\newcommand{\ea}{\end{eqnarray}}
\newcommand{\be}{\begin{equation}}

\title{HKLL bulk reconstruction for small $\Delta$}

\author{
$^a$Sinya Aoki\footnote{\tt saoki@yukawa.kyoto-u.ac.jp}\ \ and \ 
$^b$J\'anos Balog\footnote{\tt balog.janos@wigner.hu}
\\
\vskip 1ex
{\it $^a$Center for Gravitational Physics,\\
Yukawa Institute for Theoretical Physics, Kyoto University,\\
Kitashirakawa Oiwake-cho, Sakyo-Ku, Kyoto, Japan}  \\
\vskip 1ex
{\it $^b$Holographic QFT Group, Institute for Particle and Nuclear Physics,\\
Wigner Research Centre for Physics} \\
{\it H-1525 Budapest 114, P.O.B. 49, Hungary}\\
}

\received{\today} 		
\accepted{\today}	

\abstract{
We discuss the extension of the HKLL (Hamilton, Kabat, Lifschytz, and Lowe) bulk reconstruction for non-interacting
scalar fields corresponding to conformal weights $\Delta$ smaller than the
original condition $\Delta > d-1$. We give explicit formulas for the cases
$d-2<\Delta\leq d-1$ and $\Delta=d-s$ with integer $s$. 
In the latter case 
we show that smearing CFT fields over a region of the boundary
consisting of points light-like separated from the bulk point is sufficient for bulk reconstruction,
whereas in general smearing over all light-like and space-like separated points is required.
}

\begin{document}


\newcommand{\con}{\,\star\hspace{-3.7mm}\bigcirc\,}
\newcommand{\convu}{\,\star\hspace{-3.1mm}\bigcirc\,}
\newcommand{\Eps}{\Epsilon}
\newcommand{\gM}{\mathcal{M}}
\newcommand{\dD}{\mathcal{D}}
\newcommand{\gG}{\mathcal{G}}
\newcommand{\pa}{\partial}
\newcommand{\eps}{\epsilon}
\newcommand{\La}{\Lambda}
\newcommand{\De}{\Delta}
\newcommand{\nonu}{\nonumber}
\newcommand{\beq}{\begin{eqnarray}}
\newcommand{\eeq}{\end{eqnarray}}
\newcommand{\ka}{\kappa}
\newcommand{\ee}{\end{equation}}
\newcommand{\an}{\ensuremath{\alpha_0}}
\newcommand{\bn}{\ensuremath{\beta_0}}
\newcommand{\dn}{\ensuremath{\delta_0}}
\newcommand{\al}{\alpha}
\newcommand{\bm}{\begin{multline}}
\newcommand{\fm}{\end{multline}}
\newcommand{\de}{\delta}
\newcommand{\dpd}{\int {\rm d}^d p}
\newcommand{\dqd}{\int {\rm d}^d q}
\newcommand{\dxd}{\int {\rm d}^d x}
\newcommand{\dyd}{\int {\rm d}^d y}
\newcommand{\dud}{\int {\rm d}^d u}
\newcommand{\dzd}{\int {\rm d}^d z}
\newcommand{\dpp}{\int \frac{{\rm d}^d p}{p^2}}
\newcommand{\dqq}{\int \frac{{\rm d}^d q}{q^2}}


\section{Introduction and motivation}
\label{sec1}



The AdS/CFT correspondence\cite{Maldacena:1997re,Witten:1998qj}  plays a central role to investigate the holographic nature of gravity, which may give a hint for quantum gravity.
Even though much evidence has appeared after the first proposal, the fundamental mechanism why the AdS/CFT correspondence holds has not been completely understood yet. While the correspondence may be explained by the close string/open string duality, an alternative but more universal mechanism might exist because of the holographic nature of gravity.

One of the key questions one may naturally ask is how the additional dimension of the AdS emerges from CFT, which lives on the boundary of the AdS spacetime.  
An approach to this problem, called the HKLL (Hamilton, Kabat, Lifschytz, and Lowe)
bulk reconstruction,  is to relate a bulk local field operator in the AdS to CFT operators at its boundary\cite{Hamilton:2005ju,Hamilton:2006az}.
For example, let us consider a massive free scalar field operator $\Phi(X)$ with mass  squared  $m^2 =\Delta(\Delta -d)/R^2$ in the AdS with a radius $R$. Then one may define the  CFT field operator $O(t,\Omega)$ with a conformal weight $\Delta$ from $\Phi(X)$ through the BDHM relation\cite{Banks:1998dd} as
\beqa
O(x) = \lim_{\rho\to\infty} (\sinh\rho)^\Delta \Phi(t,\rho,\Omega),\quad X:=(t,\rho,\Omega),\ x:=(t,\Omega),
\label{eq:BDHM0}
\eeqa
where $\rho$ is the radial coordinate of the $d+1$ dimensional AdS with its boundary at $\rho\to\infty$, $t$ is a time coordinate,
and $\Omega$ is a $d-1$ dimensional angular variable (see section \ref{sec2}). The HKLL bulk reconstruction is the inverse mapping:
using this $O(t,\Omega)$, the bulk field can be reconstructed as
\beqa
\Phi(X) &=& \int_{\Sigma_X} dy\, K(X,y) O(y),
\label{eq:HKLL}
\eeqa
where $K(X,y)$  is a smearing function, and the integration at the boundary should be performed in a region  $\Sigma_X$ space-like separated
from the bulk point $X$. We refer to \cite{Harlow:2018fse,Kajuri:2020vxf} for recent reviews.

The result of this explicit construction can be elegantly reproduced in a somewhat abstract way\cite{Heemskerk:2012mn}. The starting point of the
abstract construction is the space-like Green function in the bulk (which vanishes if its arguments are {\it not} space-like separated). With the
help of the space-like Green function not only the free case is easily reproduced but can also be used to introduce interactions. In the original
HKLL paper (and also in this paper) the case of a free massive scalar is considered. See also \cite{Bhowmick:2019nso} for an alternative derivation based on Gel'fand-Graev-Radon transforms.
Later the reconstruction has been extended to higher spins
as well\cite{Kabat:2012hp,Kabat:2012av,Heemskerk:2012np,Kabat:2013wga,Sarkar:2014dma}. Recently an interesting connection between the bulk reconstruction and the
theory of quantum error correcting codes was pointed out\cite{Almheiri:2014lwa}.

The HKLL bulk reconstruction provides the operator to operator relation in the AdS/CFT correspondence. Recently Terashima argued under reasonable assumptions in the large $N$ limit that the relation \eqref{eq:HKLL} follows from CFT considerations {\it without} assuming the BDHM relation\cite{Terashima:2017gmc}. In other words, the BDHM relation \eqref{eq:BDHM0} is shown explicitly.
Moreover, he claimed that the integration in the space-like region $\Sigma_X$ in \eqref{eq:HKLL} can be effectively replaced by
an integration over a much smaller region $\Sigma_X^{(0)}$, which is the boundary of $\Sigma_X$
and  consists of boundary points light-like separated from $X$\cite{Terashima:2020uqu}.
(See also \cite{Terashima:2021klf}.)

Although it was not explicitly mentioned in the original papers, \eqref{eq:HKLL} holds only for $\Delta > d-1$, due to the convergence for the integral.
For applications of the AdS/CFT correspondence in the case of supersymmetric gauge theories and in particular in the prime example of the ${\cal N}=4$ SUSY U$(N)$ gauge theory in $d=4$ dimensions, 
this restriction is not essential since the conformal dimensions of physically relevant operators are typically (much) larger
than this lower bound. However, there is an other family of models often used in the AdS/CFT context, namely, the O$(N)$ vector models and their holographic
duals: higher spin theories in the bulk\cite{Klebanov:2002ja,Sezgin:2002rt}. In the most interesting $d=3$ case, for example, the simplest singlet operator
has $\Delta=1$ ($d-2$) and its square, the only relevant operator which can be used to introduce interactions, is of $\Delta=2$ ($d-1$).
These singlet scalar operators in the free $O(N)$ vector model cannot be related to the bulk operator by blindly applying \eqref{eq:HKLL}.

The case $\Delta=d-1$ was studied in \cite{Kabat:2012hp} in Poincare coordinates. It was found that in this case the support of the smearing function is the
intersection of the light-cone of the bulk point and the boundary. In \cite {DelGrosso:2019gow} the range of allowed $\Delta$ was extended to $d/2\leq\Delta\leq d-1$ by analytic continuation. Our purpose here is to find a direct derivation of the generalized HKLL formula for $\Delta$ values below the original lower bound $d-1$.

In this paper we present two results for conformal weights smaller than the lower bound mentioned above. 
We derive an extension of  the HKLL bulk reconstruction to the range $d-2< \Delta  \le d-1$, which is the first main result and is given in \eqref{calAC2}.
Our result agrees with that of \cite {DelGrosso:2019gow} (if their limit is explicitly evaluated) in the range where they overlap.
We cannot confirm Terashima's claim in general, but show that the bulk operator $\Phi(X)$ is expressed in terms of CFT operators living on
$\Sigma_X^{(0)}$ (points light-like separated from $X$ at the boundary) for the special cases $\Delta = d -s$, where $s$ is a positive integer. ($s$ is
limited by the requirement that the conformal weight satisfies the unitarity bound $\Delta>(d-2)/2$.)
This is the second main result of this paper.

\section{Review of HKLL bulk reconstruction}
\label{sec2}

In this section we review the HKLL bulk reconstruction \cite{Hamilton:2005ju,Hamilton:2006az} for a massive free scalar boson field
with conformal weight $\Delta>d-1$ in $d+1$ dimensional AdS spacetime.
This construction is very well-known, and our pupose here is to introduce our notation and conventions
and also some tools which will be needed later in the paper when we extend the validity of the construction
to smaller values of $\Delta$.

\subsection{BDHM relation}
In the Lorenttzian ${\rm AdS}_{d+1}$ space we will use
the usual global coordinates $(t,\rho,n^i)$ ($n\cdot n=1$) with the metric
\begin{equation}
{\rm d}s^2=R^2({\rm d}\rho)^2-R^2(\cosh\rho)^2({\rm d}t)^2+R^2(\sinh\rho)^2
{\rm d}n^i{\rm d}n^i,  
\label{eq:gAdS1}
\end{equation}  
where $R$ is the AdS radius.
We will denote a bulk point in ${\rm AdS}_{d+1}$ by $Y$ with global coordinates
$Y^\mu=(t,\rho,n^i)$ (with corresponding derivatives
$\partial_\mu=\partial/\partial Y^\mu$). Similarly a boundary point will be
denoted by $x$ with coordinates $x^A: (\tilde t, {\tilde n}^i)$ and derivatives
$\partial_A=\partial/\partial x^A$.
We will also use the \lq\lq flat'' coordinates $(T=Rt,y^i=R\sinh\rho \,n^i)$
and the notation  $y=\sqrt{y^iy^i}=R\sinh\rho$
for the radial coordinate. The metric in these coordinates is given by
\begin{equation}
{\rm d}s^2=-\,\frac{y^2+R^2}{R^2}({\rm d}T)^2+\left(\delta_{ij}-\frac{y^iy^j}
{y^2+R^2}\right){\rm d}y^i{\rm d}y^j.  
\label{eq:gAdS2}
\end{equation}  

In appendix \ref{appA}, we review  the complete canonical quantization of a free bulk scalar field $\Phi$ in terms of 
canonical creation and annihilation operators
${\cal A}_{n\ell{\underline m}}^+$ and ${\cal A}_{n\ell{\underline m}}$, which is given by
\begin{equation}
\Phi(t,y,\Omega)=\sum_{n\ell{\underline m}}\sqrt{\frac{{\cal N}R}{2\nu_{n\ell}}}
\Big\{u_{n\ell}(y)Y_{\ell{\underline m}}(\Omega){\cal A}_{n\ell{\underline m}}
\,{\rm e}^{-i\nu_{n\ell} t}
+u_{n\ell}(y)Y_{\ell{\underline m}}(\Omega){\cal A}_{n\ell{\underline m}}^\dagger
\,{\rm e}^{i\nu_{n\ell} t}\Big\},    
\end{equation}
where ${\cal N}$ is a normalization constant related to the free Lagrangian,
$\nu_{n\ell}=\Delta+\ell+2n$ is the eigenfrequency,
$u_{n\ell}(y)$ is the radial wave function, and $Y_{\ell{\underline m}}(\Omega)$
are hyper-spherical harmonics\footnote{We use real hyper-spherical harmonics for simplicity. This will
not be important in our analysis since we only use the hyper-spherical harmonics
$Y_{\ell {\underline 0}}$, which are real anyway.}
for the $d-1$ dimensional sphere parametrized alternatively by
the angular variables $\Omega$ or by the $d$ dimensional unit vector
$n^i$. 

The value of $\Phi$ at the middle of
(the global coordinate system of) the AdS space becomes 
\begin{equation}
\begin{split}
{\cal A}(t)=\Phi(t,0,\Omega)=\sum_n \sqrt{\frac{{\cal N}R}{2\nu_{n0}}}
\Big\{&{\rm e}^{-i\nu_{n0} t}\,(-1)^n \frac{P_n(d/2)}{n!}
{\cal N}_{n0}\frac{1}{\sqrt{\Omega_d}}{\cal A}_{n0{\underline 0}}\\
&+{\rm e}^{i\nu_{n0} t}\,(-1)^n \frac{P_n(d/2)}{n!}
{\cal N}_{n0}\frac{1}{\sqrt{\Omega_d}}{\cal A}_{n0{\underline 0}}^\dagger\Big\},
\end{split}
\end{equation}  
where $P_n(z)$ is the Pochhammer symbol, defined by
\begin{equation}
P_n(z):={\Gamma(n+z)\over \Gamma(z)}=z(z+1)\cdots (z+n-1),\qquad P_0(z)=1,
\end{equation}
$\Omega_d = \dfrac{2\pi^{d/2}}{\Gamma(d/2)} $ is a volume factor,
and the normalization constant ${\cal N}_{n\ell}$ is given by (\ref{normN}),
but it is not needed explicitly in our analysis.

With the rescaled Fock space operator,
\begin{equation}
d_n=\sqrt{\frac{{\cal N}R}{2\nu_{n0}}}\,(-1)^n \frac{P_n(d/2)}{n!}
{\cal N}_{n0}\frac{1}{\sqrt{\Omega_d}}{\cal A}_{n0{\underline 0}},
\end{equation}  
the middle-point field is expressed simply as
\begin{equation}
{\cal A}(t)={\rm e}^{-i\Delta t}D\left({\rm e}^{-2it}\right)
+{\rm e}^{i\Delta t}D_1\left({\rm e}^{2it}\right),
\label{Apm}
\end{equation}  
where formally holomorphic operators are defined by
\begin{equation}
D(z)=\sum_nd_nz^n,\qquad\qquad  D_1(z)=\sum_nd_n^\dagger z^n.  
\end{equation}  

The BDHM relation\cite{Banks:1998dd} gives
the boundary field ${\cal O}(t,\Omega)$ of conformal weight $\Delta$ as
\beqa
{\cal O}(t,\Omega):=\lim_{y\to\infty}\left(\frac{y}{R}\right)^\Delta
\Phi(t,y,\Omega)
&=&\sum_{n\ell{\underline m}} \sqrt{\frac{{\cal N}R}{2\nu_{n\ell}}}
\Big\{{\rm e}^{-i\nu_{n\ell} t}\,\frac{P_n(1+\alpha)}{n!}
{\cal N}_{n\ell}Y_{\ell{\underline m}}(\Omega){\cal A}_{n\ell{\underline m}}\nn \\
&+&{\rm e}^{i\nu_{n\ell} t}\,\frac{P_n(1+\alpha)}{n!}{\cal N}_{n\ell}Y_{\ell{\underline m}}(\Omega){\cal A}_{n\ell{\underline m}}^\dagger \Big\},
\eeqa  
where $\alpha:=\Delta-d/2$ (see appendix~\ref{appA}).
It is clear that $O(t,\Omega)$ in the above expression is not a canonical field operator, 
since it does not satisfy the canonical commutation relation $[O(t,\Omega), \partial_t O(t,\Omega^\prime)]=i\delta(\Omega-\Omega^\prime)$.

An integration over the angular variables simplifies the above formula as
\beqa
{\cal C}(t)&:=&\int{\rm d}\Omega\, {\cal O}(t,\Omega) = e^{-i\Delta t} B(-e^{-2it}) + e^{i\Delta t} B_1(-e^{2it}), 
\label{Cpm}
\eeqa
where an other pair of formally holomorphic operators is given by
\begin{equation}
B(z)=\sum_nb_nz^n,\qquad\qquad  B_1(z)=\sum_nb_n^\dagger z^n 
\end{equation}  
in terms of   Fock space operators  rescaled differently from $d_n$ as
\begin{equation}
b_n=\sqrt{\frac{{\cal N}R}{2\nu_{n0}}}\,(-1)^n \frac{P_n(1+\alpha)}{n!}
{\cal N}_{n0}\sqrt{\Omega_d}{\cal A}_{n0{\underline 0}}
= \Omega_d\,\frac{P_n(1+\alpha)}{P_n(d/2)} d_n .
\label{ratio}
\end{equation} 

\subsection{Bulk-boundary mapping}

Following HKLL\cite{Hamilton:2005ju,Hamilton:2006az}, we relate the holomorphic functions $D$ and $B$ as
\begin{equation}
D(w)=\sum_n\frac{1}{\Omega_d}\frac{P_n(d/2)}{P_n(1+\alpha)}w^n\frac{1}{2\pi i}
\oint \frac{{\rm d}z}{z^{n+1}}B(z),
\end{equation}  
which, by reversing the order of summation and integration, is rewritten as 
\beqa
D(w)&=&\frac{1}{2\pi i\Omega_d}\oint\frac{{\rm d}z}{z}B(z)\sum_n
\frac{P_n(d/2)}{P_n(1+\alpha)}\left(\frac{w}{z}\right)^n
=\frac{1}{2\pi i\Omega_d}\oint\frac{{\rm d}z}{z}B(z)
{}_2F_1(1,d/2;1+\alpha;w/z)\nn\\
&=&\frac{1}{2\pi i\Omega_d}\oint\frac{{\rm d}z}{z}B(wz)
{}_2F_1(1,d/2;1+\alpha;1/z).
\label{Dw1}
\eeqa
The integration contour in the last formula must lie outside the unit circle
for the sum defining the hypergeometric function
to  be convergent.

In this paper, for simplicity\footnote{Similar results hold also for even
$d$, but some of the formulas receive logarithmic corrections\cite{Hamilton:2005ju,Hamilton:2006az}.},
we mainly (except in subsection 4.2) consider the case $d$ odd. The derivation of the
explicit form
of the linear relation between the bulk field \lq\lq at the middle'' and
the integrated boundary field found by HKLL is reproduced in
appendix \ref{appC}. Although it was not emphasized in the original HKLL paper\cite{Hamilton:2006az},
this derivation is valid for the range
\begin{equation}
\Delta>d-1.
\label{case}
\end{equation}
only. The result is given by
\begin{equation}
{\cal A}(t)=\xi
\int_{t-\pi/2}^{t+\pi/2}{\rm d}u [2\cos(t-u)]^{\Delta-d}{\cal C}(u),
\label{calAC}
\end{equation}
where the overall constant is
\begin{equation}
\xi=\frac{1}{\pi\Omega_d}\frac{\Gamma(1-d/2)\Gamma(1+\alpha)}
{\Gamma(\Delta-d+1)}.
\label{eq:xi}
\end{equation}  
We can see that the HKLL result (\ref{calAC}) is valid for the range
(\ref{case}) only, because for $\Delta\leq d-1$ this integral is divergent.
In the next section and appendix \ref{appD}, we extend the calculation for
$\Delta>d-2$ and consider the most interesting special case $\Delta=d-1$ in
detail.


We finish the review of the HKLL construction by transforming the result,
calculated above for the ``middle'' of the AdS space,
to an arbitrary point in AdS space.
The result (\ref{calAC}) for the ``middle'' point
$Y_o=(t=0, \rho=0,\Omega)$ is rewritten as
\begin{equation}
\Phi(Y_o)=\int{\cal D}x\, {\cal K}(x){\cal O}(x),
\end{equation}
where
\begin{equation}
x=(\tilde t,\tilde \Omega),\quad 
{\cal D}x={\rm d}\tilde t{\rm d}\tilde\Omega,\qquad
{\cal K}(x)=\xi(2\cos\tilde t)^{\Delta-d}\Theta\left(\frac{\pi}{2}-\tilde t\right)
\Theta\left(\tilde t+\frac{\pi}{2}\right)
\end{equation}
with the step function $\Theta$.

In what follows we will make use the symmetry properties of the solution
and use the notations introduced in appendix \ref{appB}.
Applying the Hilbert space isometry action to both sides of the equation, 
$\Phi$ for a generic bulk point $Y=g^{-1} Y_o$ is represented as 
\begin{equation}
\Phi(Y=g^{-1} Y_o)=\int{\cal D}x\, {\cal K}(x)[J(g^{-1},x)]^\Delta{\cal O}(g^{-1}x)
=\int{\cal D}y\, {\cal K}(gy)[J(g,y)]^{d-\Delta}{\cal O}(y),
\label{eq:cond_Phi}
\end{equation}
where (\ref{Jid}) is used for the second equality.
The solution to the above equation is given by
\begin{equation}
\Phi(Y=g^{-1} Y_o)=\int{\cal D}x\, I^{\Delta-d}(Y,x) T(Y,x){\cal O}(x),
\label{eq:sol_Phi}
\end{equation}
where $I$ and $T$  have to satisfy
\beqa
I(gY,gx)&=&J(g,x)I(Y,x),\quad I(Y_o,x)=2\cos\tilde t, 
\label{Ifun}
\\
T(gY,gx)&=&T(Y,x),\quad
T(Y_o,x)=\xi\Theta\left(\tilde t+\frac{\pi}{2}\right)\Theta\left(\frac{\pi}{2}-\tilde t\right).
\label{Tfun}
\eeqa
Now it is easy to see that \eqref{eq:sol_Phi} satisfies \eqref{eq:cond_Phi} since
\beqa
\Phi(Y=g^{-1} Y_o)&=& \int{\cal D}y\, J^{d-\Delta}(g,y) I^{\Delta-d}(Y_o,gy) T(Y_o,gy){\cal O}(y)\nn \\
&=&  \int{\cal D}y\,  J^{d-\Delta}(g,y) {\cal K}(gy) O(y).
\eeqa
$I$ and $T$ for $Y=(t,\rho,n^i)$ and $x=(\tilde t,\tilde n^i)$ are explicitly constructed in appendix \ref{appE}:
\beqa
I(Y,x)&=&2[\cosh\rho \cos(t-\tilde t)-\sinh\rho \, n\cdot {\tilde n}], \quad
T(Y,x)=\xi\Theta(X_1)\Theta(X_2),
\eeqa
where $X_1=\tilde t-T_1$, $X_2=T_2-\tilde t$, and $T_{1,2}$ are defined in (\ref{T1}) and (\ref{T2}). 
Geometrically, if $X_{1}(Y,x)=0$ or $X_{2}(Y,x)=0$, $Y$ and $x$ can be
connected by a past or future oriented light-like geodesic, respectively.
Thus  $\Theta(X_1)\Theta(X_2)$ is only
non-vanishing if $T_1<\tilde t<T_2$,
which means that $Y$ and $x$ can be connected by a space-like geodesic. This
last observation leads to the introduction of the space-like Green function,
which is useful to introduce interactions in the bulk. (See \cite{Harlow:2018fse} for a review.)

\section{Bulk reconstruction for the range $d-2<\Delta\leq d-1$}
\label{sec3}

We have seen that the derivation of the HKLL
formula is only valid for the range (\ref{case}).
(The a priori lower limit for a scalar field is $\Delta > (d-2)/2$, which is smaller.)
Here we extend the possible range to
\begin{equation}
\Delta>d-2.
\label{case2}
\end{equation}
Our starting point is the last line of (\ref{Dw1}) and the identity (\ref{hyp}).
We note that this hypergeometric identity is valid for odd $d$ and
$\Delta\not={\rm integer}$. This last requirement is only temporary and later
we extend the results (by taking limits) to integer $\Delta$, too.

To circumvent the restriction (\ref{case}), we rewrite (\ref{Dw1}) by adding
and subtracting $B(w)$ under the integral as
\beqa
D(w)
&=&\frac{B(w)}{2\pi i\Omega_d}\oint\frac{{\rm d}z}{z}\,
{}_2F_1(1,d/2;1+\alpha;1/z)\nn \\
&+&\frac{1}{2\pi i\Omega_d}\oint\frac{{\rm d}z}{z}[B(wz)-B(w)]\,
{}_2F_1(1,d/2;1+\alpha;1/z).
\eeqa
Using this form, the manipulations in appendix \ref{appC}
remain valid for the extended range
$\Delta>d-2$ and we obtain
\begin{equation}
D(w)=\frac{B(w)}{\Omega_d}+\xi
\int_{-\pi/2}^{\pi/2}{\rm d}u\, {\rm e}^{-iu\Delta}[2\cos(u)]^{\Delta-d}
\{B(-w{\rm e}^{-2iu})-B(w)\},
\label{Dw}
\end{equation}
where singularities near $u=\pm {\pi\over 2}$ of the integrand become integrable for $\Delta > d-2$ thanks to the subtraction of $B(w)$.

Employing the above expression for $D(w)$ and a similar one for $D_1(w)$ (see appendix \ref{appD} for
the details of the derivation), we obtain one of our main results in this paper:
\beqa
{\cal A}(t)&=&\frac{\eta}{2\Omega_d}[{\cal C}(t-\pi/2)+{\cal C}(t+\pi/2)]
+\xi\int_{t-\pi/2}^t{\rm d}u [2\cos(u-t)]^{\Delta-d}\{{\cal C}(u)
-{\cal C}(t-\pi/2)\}\nn \\
&+&\xi\int_t^{t+\pi/2}{\rm d}u [2\cos(u-t)]^{\Delta-d}\{{\cal C}(u)
-{\cal C}(t+\pi/2)\},
\label{calAC2}
\eeqa
which is valid for the extended range (\ref{case2}).
Here
\begin{equation}
\eta=\frac{\Gamma(1-d/2)\Gamma(1+\alpha)}{\Gamma^2(1+\frac{\Delta-d}{2})}.
\end{equation}
Our explicit derivation confirms the result found in \cite{DelGrosso:2019gow} (if the limit is explicitly evaluated) at least in their overlapping
range of validity.
For the original range, $\Delta>d-1$, 
\eqref{calAC2} gives back the original
HKLL result (\ref{calAC}),
since the subtracted terms, which now can be integrated separately by using the
identity
\begin{equation}
\int_0^{\pi/2}{\rm d}u(2\cos u)^A
=\frac{\pi}{2}\,\frac{\Gamma(1+A)}{\Gamma^2(1+A/2)},\qquad\quad A>-1,
\end{equation}
exactly cancel the first term.

An interesting special case is obtained if we take the limit $\Delta\to d-1$.
In this limit,  the integrals do not contribute as $\xi=0$, and 
$\eta$ simplifies to $\eta=(-1)^{\Delta/2}$.
We thus obtain
\begin{equation}
{\cal A}(t)=\xi_o [{\cal C}(t-\pi/2)+{\cal C}(t+\pi/2)],\quad \xi_o :=  \frac{(-1)^{\Delta/2}}{2\Omega_d},
\label{specdm1}
\end{equation}
which means that  the bulk field at the middle point in the global AdS is expressed in terms of the CFT field values only at boundary points
connected to the middle point by light-like geodesics.   
This is the other main result in this paper, which  is in agreement with  the result in \cite{Kabat:2012hp} and
confirms the claim in \cite{Terashima:2021klf} for the special case $\Delta=d-1$.
We will consider this interesting case and its generalization to  $\Delta = d- s$ with an integer $s$ in the next section.

It is also straightforward to extend the range to $\Delta > d-3$, by rewriting (\ref{Dw1}) as
\beqa
D(w)
&=&\frac{1}{2\pi i\Omega_d}\oint\frac{{\rm d}z}{z}\left[B(w) + B^\prime(w)w(z-1)\right]\,
{}_2F_1(1,d/2;1+\alpha;1/z)\nn \\
&+&\frac{1}{2\pi i\Omega_d}\oint\frac{{\rm d}z}{z}[B(wz)-B(w)-B^\prime(w)w(z-1)]\,
{}_2F_1(1,d/2;1+\alpha;1/z),
\eeqa
but we do not pursue this direction further in this paper.



\section{Bulk reconstruction for $\Delta=d-s$ with an integer $s$}
\label{sec4}

In this section we consider the special cases $\Delta=d-s$ with integer $s < (d+2)/2$ satisfying 
the lower bound, $\Delta > (d-2)/2$.
For these special cases we have found a simpler derivation of the bulk reconstruction formulas, in particular for \eqref{specdm1} with odd $d$,
without using the limiting procedure starting from integrals like \eqref{calAC2}. 
Interestingly the bulk field operator at the middle point can be expressed in terms of CFT field operators and their $t$ derivatives
only at boundary points light-like separated from the middle point. This is shown by \eqref{eq:d-odd} and \eqref{eq:d-even},
which are also one of our main results in this paper.
For even $d$, we can derive similar results, which however also contain a derivative with respect to $\Delta$.

From \eqref{ratio} we see that the bulk fleld can be written in terms of boundary operators $b_n$ and $b_n^\dagger$ as
\beqa
{\cal A}(t) &=& {1\over \Omega_n}\sum_n X_n^\Delta \left\{e^{-i(\Delta+2n) t} b_n + e^{i(\Delta+2n) t} b_n^\dagger \right\},
\quad X_{n}^{\Delta} := {P_n(\frac{d}{2})\over P_n(\alpha+1)}.
\eeqa
On the other hand we have
\beqa
C_\pm (t) &:=& {\cal C} \left(t+{\pi\over 2}\right) \pm {\cal C}\left(t-{\pi\over 2}\right)\nn \\
&=&\left(e^{-i\Delta {\pi\over 2}}\pm e^{i\Delta {\pi\over 2}}\right)
\sum_n  \left\{e^{-i(\Delta+2n) t} b_n \pm e^{i(\Delta+2n) t} b_n^\dagger \right\}.
\eeqa
Thus $C_\pm (t) =0$ if $\Delta$ is an odd/even integer. 

\subsection{Results for $\Delta=d-s$ with odd $d$}

As a warmup, we first give a much simpler derivation of \eqref{specdm1} for $\Delta=d-1$.
Since $X_n^{\Delta}=1$ in this case,
we have
\beqa
{\cal A}(t)&=&{1\over\Omega_d}\sum_n \left\{e^{-i(\Delta+2n) t} b_n + e^{i(\Delta+2n) t} b_n^\dagger \right\}
={(-1)^{\Delta/2}\over 2\Omega_d} C_+(t),\quad \Delta=d-1,
\eeqa
which reproduces \eqref{specdm1}, because $C_+(t)= 2{\cal C}(t\pm{\pi\over 2})$.

For $\Delta=d-2$, since $X_n^{\Delta} = (\Delta+2n)/(d-2)$, we obtain 
\beqa
{\cal A}(t) ={1\over \Omega_d}\sum_n X_n^\Delta\left\{e^{-i(\Delta+2n) t} b_n + e^{i(\Delta+2n) t} b_n^\dagger \right\}
= -{(-1)^{\Delta-1\over 2}\over 2(d-2)\Omega_d}{\partial\over \partial t} C_-(t),
\eeqa
where in this case $C_-(t)= \pm 2{\cal C}(t\pm{\pi\over 2})$.

For general $\Delta = d-s$ with $s < (d+2)/2$, we have
\beqa
X_n^{d-(2\ell +1)} &=&X_n^{d-1} {\prod_{k=1}^{\ell} (\Delta_n+2k-1)(\Delta_n-2k+1)\over \prod_{k=1}^{2\ell}(d-2k)}, \quad
X_n^{d-1}=1,\\
X_n^{d-(2\ell +2)}&=&X_n^{d-2} {\prod_{k=1}^{\ell} (\Delta_n+2k)(\Delta_n-2k)\over \prod_{k=2}^{2\ell+1}(d-2k)}, \quad
X_n^{d-2}={\Delta_n\over d-2}
\eeqa
for $\ell=1,2,\cdots$, where $\Delta_n := \Delta + 2n$.
We thus obtain
\beqa
{\cal A}(t) &=&{(-1)^{d-1\over 2}\over 2\Omega_d}{\displaystyle\prod_{k=1}^\ell \left\{ \dfrac{\partial^2}{\partial t^2} +(2k-1)^2\right\} \over \prod_{k=1}^{2\ell} (d-2k)} C_+(t),
\label{eq:d-odd}
\eeqa
for $\Delta=d-(2\ell+1)$, where $C_+(t) =2{\cal C}(t\pm {\pi\over 2})$, while
\beqa
{\cal A}(t) &=&{(-1)^{d-1\over 2}\over 2\Omega_d}{\displaystyle\prod_{k=1}^{\ell} \left\{ \dfrac{\partial^2}{\partial t^2} +4k^2\right\} \over \prod_{k=1}^{2\ell+1} (d-2k)} {\partial\over \partial t} C_-(t),
\label{eq:d-even}
\eeqa
for $\Delta =d-2(\ell+1)$,  where $C_-(t) =\pm 2{\cal C}(t\pm {\pi\over 2})$.
(\ref{eq:d-odd}) and (\ref{eq:d-even}) cover all cases $\Delta = d- s$ for odd $d$.  

\subsection{Results for $\Delta=d-s$ with even $d$}

For an even dimension $d$, the bulk field operator becomes
\beqa
{\cal A}(t)&=& {(-1)^\ell\over \Omega_d} {\displaystyle\prod_{k=1}^\ell \left\{\dfrac{\partial^2}{\partial t^2} +(2k-1)^2 \right\} \over \prod_{k=1}^{2\ell} (d-2k)}
\sum_n  \left\{e^{-i(\Delta+2n) t} b_n + e^{i(\Delta+2n) t} b_n^\dagger \right\}
\eeqa
for $\Delta = d-(2\ell+1)$, while
\beqa
{\cal A}(t)&=& {(-1)^{\ell}\over \Omega_d}
{\displaystyle\prod_{k=1}^{\ell} \left\{ \dfrac{\partial^2}{\partial t^2} +4k^2\right\} \over \prod_{k=1}^{2\ell+1} (d-2k)} {i\partial\over \partial t}
\sum_n  \left\{e^{-i(\Delta+2n) t} b_n - e^{i(\Delta+2n) t} b_n^\dagger \right\}
\eeqa
for $\Delta = d-2(\ell+1)$.
On the other hand, the boundary field operators satisfy
\beqa
\left. {\partial\over \partial\Delta} C_+(t) \right\vert_{\Delta=d-(2\ell+1)} &=& \pi (-1)^{d/2-\ell} \sum_n  \left\{e^{-i(\Delta+2n) t} b_n + e^{i(\Delta+2n) t} b_n^\dagger \right\}, \\
\left. {\partial\over \partial\Delta} C_-(t) \right\vert_{\Delta=d-2(\ell+1)} &=&(-i) \pi (-1)^{d/2-\ell-1} \sum_n  \left\{e^{-i(\Delta+2n) t} b_n - e^{i(\Delta+2n) t} b_n^\dagger \right\}.
\eeqa
Combining these, we obtain
\beqa
{\cal A}(t) &=& {(-1)^{d/2}\over \pi\Omega_d}  {\displaystyle\prod_{k=1}^\ell \left\{\dfrac{\partial^2}{ \partial t^2} +(2k-1)^2 \right\} \over \prod_{k=1}^{2\ell} (d-2k)}
\left. {\partial\over \partial\Delta} C_+(t) \right\vert_{\Delta=d-(2\ell+1)},
\label{eq:ed-even}\\
{\cal A}(t) &=&  {(-1)^{d/2}\over \pi\Omega_d} {\displaystyle\prod_{k=1}^{\ell} \left\{ \dfrac{\partial^2}{ \partial t^2} +4k^2\right\} \over \prod_{k=1}^{2\ell+1} (d-2k)} {\partial\over \partial t} \left. {\partial\over \partial\Delta} C_-(t) \right\vert_{\Delta=d-2(\ell+1)} .
\label{eq:ed-odd}
\eeqa

For $\Delta=d-1,d-2$, for example, we have
\beqa
{\cal A}(t) &=& {(-1)^{d/2}\over \pi\Omega_d}  \left. {\partial\over \partial\Delta} C_+(t) \right\vert_{\Delta=d-1},\quad
{\cal A}(t) =  {(-1)^{d/2}\over (d-2)\pi\Omega_d}{\partial\over \partial t}  \left. {\partial\over \partial\Delta} C_-(t) \right\vert_{\Delta=d-2} .
\eeqa

\section{Bulk reconstruction at generic points for small integer $\Delta$}

In this section we derive the bulk field operator at generic points for $\Delta=d-1$ and $\Delta=d-2$ with odd $d$, 
along the same logic we used in section \ref{sec2} for the (\ref{case}) case.

\subsection{Bulk reconstruction for $\Delta=d-1$ with odd $d$ at generic bulk points}

For the middle point $Y_o$, we write
\begin{equation}
\Phi(Y_o)= \int{\cal D}x\, k(x){\cal O}(x),\qquad\quad k(x):=\xi_o[\delta_o(\tilde t+\pi/2)
+\delta_o(\tilde t-\pi/2)],
\end{equation}
where $\delta_o$ is the standard delta function of one argument.
Making the isometry transformation in the Hilbert space as before, we obtain
\begin{equation}
\Phi(g^{-1}Y_o)=\int{\cal D}y J^d(g,y)J^\Delta(g^{-1},gy) k(gy){\cal O}(y)
=\int{\cal D}y \,k(gy) J(g,y)
{\cal O}(y),
\end{equation}
which is solved by
\begin{equation}
\Phi(Y=g^{-1} Y_o)=\int{\cal D}x\,D(Y,x){\cal O}(x),
\end{equation}
where $D(Y,x)$  has to satisfy
\begin{equation}
D(gY,gx)=\frac{D(Y,x)}{J(g,x)},\quad
D(Y_o,x)=k(x),
\end{equation}
since, with this definition,
\beqa
\Phi(g^{-1}Y_o)&=&\int{\cal D}x\,D(g^{-1}Y_o,x){\cal O}(x)
=\int{\cal D}x\,J(g,x)\,k(gx){\cal O}(x).
\eeqa

The kernel function $D(Y,x)$ is constructed explicitly in appendix \ref{appF}
and is given by
\begin{equation}
D(Y,x)=\frac{\xi_o}{{\cal R}(Y,x)}[\delta_o(X_1)+\delta_o(X_2)],
\label{kernelD}
\end{equation}
where
\begin{equation}
{\cal R}(Y,x)=\cosh\rho\cos\Psi=\sqrt{\cosh^2\rho-\sinh^2\rho( n\cdot
{\tilde n})^2},\qquad\quad {\cal R}(Y_o,x)=1.
\label{eq:defR}
\end{equation}
The final result for the bulk reconstruction for $\Delta=d-1$ with odd $d$ is
\begin{equation}
\Phi(Y)=\xi_o\int{\rm d}\tilde\Omega\frac{1}{{\cal R}(Y,x)}[{\cal O}(T_1,
\tilde\Omega)+{\cal O}(T_2,\tilde\Omega)],
\label{eq:generic_d-1}
\end{equation}
which again shows 
that the field operator at  a generic bulk point is reconstructed 
from operators having support only on boundary points light-like separated from the bulk point. 

Although the BDHM relation \cite{Banks:1998dd} was our starting point in the construction, it is by far not
obvious that the representation \eqref{eq:generic_d-1} reproduces this relation. It is a nice check on our results that, 
as explicitly shown in appendix \ref{appG},  $\Phi(Y)$ in \eqref{eq:generic_d-1} for $Y=(t,\rho,\Omega)$ does satisfy 
the BDHM relation
\beqa
\lim_{\rho\to\infty} (\sinh\rho)^\Delta \Phi(t,\rho,\Omega) = {\cal O}(t,\Omega).
\label{eq:BDHM}
\eeqa

\subsection{Bulk reconstruction for $\Delta=d-2$ with odd $d$ at generic bulk points}

For this special case the bulk reconstruction at the origin can be written in
a symmetric way as
\begin{equation}
{\cal A}(t)=-\tilde\xi_o\frac{\partial}{\partial t}[
{\cal C}(t+\pi/2)-{\cal C}(t-\pi/2)],
\qquad\quad
\tilde\xi_o=\frac{(-1)^{\frac{\Delta-1}{2}}}{2(d-2)\Omega_d}.
\label{dm2}
\end{equation}
To extend the result to an arbitrary bulk point we can proceed
analogously to the $\Delta=d-1$ case.

In a more compact notation (\ref{dm2}) for the middle point $Y_o$
can be written as
\begin{equation}
\Phi(Y_o)=\int{\cal D}x k_2(x){\cal O}(x),\qquad\quad
k_2(x)=-\tilde\xi_o[\delta^\prime_o(\tilde t+\pi/2)
-\delta^\prime_o(\tilde t-\pi/2)].
\end{equation}
Here $\delta^\prime_o$ is the derivative of the delta function.
Making the isometry transformation in the Hilbert space, we find
\begin{equation}
\begin{split}
\Phi(g^{-1}Y_o)&=\int{\cal D}x\, k_2(x)J^\Delta(g^{-1},x) {\cal O}(g^{-1}x)
=\int{\cal D}y \,k_2(gy) J^2(g,y) {\cal O}(y).
\end{split}
\label{dm2Yo}
\end{equation}
Motivated by the $\Delta=d-1$ result we now take the ansatz $\Phi(Y)=\int{\cal D}x\,D_2(Y,x){\cal O}(x)$,
where we require that $D_2(Y,x)$ satisfies
\begin{equation}
D_2(gY,gx)=\frac{D(Y,x)}{J^2(g,x)},\quad
D_2(Y_o,x)=k_2(x).
\label{cond2}
\end{equation}
We can now verify that (\ref{dm2Yo}) holds:
\begin{equation}
\Phi(g^{-1}Y_o)=\int{\cal D}x\,D_2(g^{-1}Y_o,x){\cal O}(x)
=\int{\cal D}x\,J^2(g,x)\,k_2(gx){\cal O}(x).
\end{equation}

The derivation of the solution of (\ref{cond2}) is given in
appendix \ref{appF}: 
\begin{equation}
D_2(Y,x)=-\frac{\tilde\xi_o}{{\cal R}^2(Y,x)}
\left\{[\delta^\prime_o(X_1)+\delta^\prime_o(X_2)]
 -\tan\Psi[\delta_o(X_1)+\delta_o(X_2)]\right\}.
\end{equation}





\section{Conclusions and discussion}
\label{concl}

In this paper we extended the applicability of the HKLL bulk reconstruction for non-interacting scalar theories,
which was restricted to $\Delta > d-1$, to smaller conformal weights $\Delta$ of the boundary CFT in the range $d-2 <\Delta \le d-1$.
The explicit formula is given in \eqref{calAC2}.
In addition, we have derived a simple formula for $\Delta=d-s$ with positive integer $s$,
which (for these special cases) confirms Terashima's claim that
a field at a point $X$ in the AdS bulk can be reconstructed from CFT fields smeared only over boundary points 
connected to $X$ by light-like geodesic curves\cite{Terashima:2020uqu,Terashima:2021klf}.
 
Results in this paper enable us to apply the HKLL bulk reconstruction to $O(N)$ vector models,
which are expected to be dual to higher spin theories\cite{Klebanov:2002ja,Sezgin:2002rt}.
Moreover, explicit demonstration of Terashima's claim, even though only for the above special cases,
may bring new insights \cite{Terashima:2021klf} to the sub-region duality and its relation to quantum error corrections\cite{Almheiri:2014lwa}.
 
It would be interesting to generalize the Green function method so that it covers the extended range and to reproduce \eqref{calAC2} with this technique.
We hope that this will enable us to introduce interactions systematically.

\vspace{5ex}
%

\section*{Acknowledgements}
We thank Dr. S. Terashima for useful discussions and for partially checking our results.
J.B. acknowledges support from the International Research Unit of Quantum Information (QIU) of Kyoto University Research Coordination Alliance,
and also would like to thank the Yukawa Institute for Theoretical Physics at Kyoto University, where most of this work has been carried out, for support and hospitality by the long term visitor program. 
This work has been supported in part by the NKFIH grant K134946, 
by the Grant-in-Aid of the Japanese Ministry of Education, Sciences and Technology, Sports and Culture (MEXT) for Scientific Research (Nos.~JP16H03978,  JP18H05236).

\par\bigskip

\appendix

\section{Canonical quantization of the free scalar field}
\label{appA}

Quantization in a general curved background is difficult, but it is
straightforward if there exists a global time $t$ and the metric has a
form,
\begin{equation}
{\rm d}s^2=-{\cal H}({\rm d}t)^2+g_{ij}{\rm d}x^i{\rm d}x^j,
\end{equation}  
where $\{x^1,\dots,x^d\}$ are the space coordinates, and 
both ${\cal H}$ and $g_{ij}$ are time-independent.
In such a background the Lagrangian of a (dimensionless) free scalar $\Phi$ is defined as  
\begin{equation}
L=\frac{1}{{2\cal N}}\dxd\frac{\sqrt{-g}}{{\cal H}}[{\dot\Phi}^2-\Phi K\Phi],  
\label{lag}
\end{equation}  
where ${\cal N}=B^{d-1}$, $B$ is a parameter of length dimension,
\begin{equation}
K=-\frac{{\cal H}}{\sqrt{-g}}\partial_i\left(\sqrt{-g}g^{ij}\partial_j\right)
+{\cal H}m^2,  
\end{equation}  
and $m^2$ is a parameter of dimension mass squared. 
The consistency of the quantization procedure requires that the operator $K$
is self-adjoint such that
$\langle f_1\vert K f_2\rangle = \langle K f_1\vert f_2\rangle$
for any two functions $f_1$, $f_2$ in the domain of definition of $K$,
where the scalar product of two functions is defined with the measure  $\sqrt{-g}/{\cal H}$ as
\begin{equation}
\langle f_1\vert f_2\rangle=\dxd\frac{\sqrt{-g}}{{\cal H}}f_1\,f_2.  
\end{equation} 
The Euler-Lagrange equations following from the Lagrangian (\ref{lag}) can be
written in a covariant form as
\begin{equation}
\frac{1}{\sqrt{-g}}\partial_\mu\left(\sqrt{-g}g^{\mu\nu}\partial_\nu\Phi\right)
=m^2\Phi.
\label{EoM}
\end{equation}  

We will expand the free field in terms of eigenfunctions of $K$ satisfying
\begin{equation}
K\psi_a=\omega_a^2\psi_a,\qquad\qquad \langle\psi_a\vert\psi_b\rangle
=\delta_{ab}, 
\end{equation}  
where the frequencies $\omega_a$ are all real because $K$ (for large enough $m^2$)
is positive self-adjoint. Writing the field as
\begin{equation}
\Phi(x)=\sum_aQ_a\psi_a(x)
\end{equation}  
the Lagrangian becomes 
\begin{equation}
L=\frac{1}{2{\cal N}}\sum_a[{\dot Q_a}^2-\omega_a^2{Q_a}^2].  
\end{equation}  
A complete set of solutions to the equations of motion (\ref{EoM}) is
$\{f_a(t,x)\}$, $\{f_a^*(t,x)\}$,    
where $f_a(t,x)={\rm e}^{-i\omega_at}\psi_a(x)$, so that
the general solution is expanded in terms of constant amplitudes $\{\beta_a\}$ as
\begin{equation}
\Phi(t,x)=\sum_a[f_a(t,x)\beta_a+f_a^*(t,x)\beta_a^*].
\end{equation}  
 
We introduce canonical momentum variables and the Hamiltonian of the system as
\begin{equation}
H=\frac{1}{2}\sum_a \left({\cal N}p_a^2+\frac{\omega^2_a}{{\cal N}}Q_a^2\right)  
=\frac{2}{{\cal N}}\sum_a\omega_a^2\beta^*_a\beta_a, \quad p_a:=\frac{1}{{\cal N}}{\dot Q}_a,
\end{equation} 
then 
we promote the canonical variables $p_a$, $Q_a$ to operators satisfying
$ [p_a,Q_b]=-i\delta_{ab}$.
The quantized amplitudes become
\begin{equation}
\beta_a=\sqrt{\frac{{\cal N}}{2\omega_a}}{\cal A}_a,\qquad\quad  
\beta^+_a=\sqrt{\frac{{\cal N}}{2\omega_a}}{\cal A}^\dagger_a,  
\end{equation}   
where ${\cal A}_a$, ${\cal A}_a^\dagger$ are operators in a Fock space such that
$ [{\cal A}_a,{\cal A}_b^\dagger]=\delta_{ab}$ and ${\cal A}_a\vert0\rangle=0$, and
the corresponding quantum Hamiltonian becomes
\begin{equation}
H=E_0+\sum_a\omega_a{\cal A}_a^\dagger{\cal A}_a,  
\end{equation}  
where $E_0$ is the vacuum energy and $\{\omega_a\}$ is the spectrum of
1-particle states in the Fock space.
Finally the canonical quantum field operator is expanded as
\begin{equation}
\Phi(t,x)=\sum_a\sqrt{\frac{{\cal N}}{2\omega_a}}\left[f_a(t,x){\cal A}_a+
f_a^*(t,x){\cal A}_a^\dagger\right].  
\end{equation}  

\subsection{Radial quantization}

We will apply the above quantization scheme to the global AdS space.
For the metric \eqref{eq:gAdS2}, we have
\begin{equation}
{\cal H}=\frac{y^2+R^2}{R^2},\qquad\quad \sqrt{-g}=1,\qquad\quad
g^{ij}=\delta_{ij}+\frac{y^iy^j}{R^2},  
\end{equation}  
and the operator $K$ becomes
\begin{equation}
K={\cal H}\left\{\frac{L^2}{y^2}-\left(\frac{1}{y^2}+\frac{1}{R^2}\right)
({\cal D}^2+d{\cal D})+\frac{2}{y^2}{\cal D}+m^2\right\},  
\end{equation}  
where
\begin{equation}
L^2:=-\frac{1}{2}L_{ij}L_{ij},\qquad
L_{ij}:=y^i\frac{\partial}{\partial y^j}-y^j\frac{\partial}{\partial y^i}, \qquad
{\cal D}:=y^i\frac{\partial}{\partial y^i}.
\end{equation}  
Eigenfunctions of the Casimir operator $L^2$ are hyper-spherical
harmonics $Y_{\ell \underline{m}}(\Omega)$, where $\Omega$ are the angular
variables ($n^i$). The spectrum is given by
\begin{equation}
L^2\,Y_{\ell \underline{m}}(\Omega)=\ell(\ell+d-2)\,Y_{\ell \underline{m}}(\Omega),  
\end{equation}  
where $\ell=0,1,\dots$ and $\underline{m}$ is a multi-index. The hyper-spherical
harmonics in a real basis are normalized  to 
\begin{equation}
\int{\rm d}\Omega\,Y_{\ell^\prime \underline{m}^\prime}(\Omega)
Y_{\ell \underline{m}}(\Omega)=\delta_{\ell^\prime \ell}
\delta_{\underline{m}^\prime\underline{m}},  
\end{equation}
where ${\rm d}\Omega$ is the measure of the angular integration, $\dyd=\int_0^\infty{\rm d}y\,y^{d-1}{\rm d}\Omega$.
Using the ansatz $\psi_{\ell\underline{m}}(y,\Omega)=u_\ell(y)Y_{\ell\underline{m}}(\Omega)$ for eigenfunctions, 
the radial functions $u_\ell(y)$ must satisfy the differential equation
\begin{equation}
{\cal H}\left\{\frac{\ell(\ell+d-2)}{y^2}-\left(\frac{1}{y^2}+\frac{1}{R^2}
\right)({\cal D}^2+d{\cal D})+\frac{2}{y^2}{\cal D}+m^2\right\}u_\ell(y)
=\omega^2_\ell \,u_\ell(y), 
\label{rad1}
\end{equation}
where now ${\cal D}=y\frac{{\rm d}}{{\rm d}y}$, 
and the radial scalar product is defined by
\begin{equation}
\langle u^{(1)}_\ell\vert u^{(2)}_\ell\rangle=
\int_0^\infty{\rm d}y\,y^{d-1}\frac{R^2}{y^2+R^2}u^{(1)}_\ell(y)\,u^{(2)}_\ell(y).  
\label{scalar}
\end{equation}  
Introducing dimensionless quantities
\begin{equation}
\mu=mR,\qquad\quad \nu_\ell=\omega_\ell R,\qquad\quad
\xi=\frac{y^2}{y^2+R^2},\qquad\quad 1-\xi=\frac{R^2}{y^2+R^2},  
\end{equation}  
(\ref{rad1}) becomes
\begin{equation}
K^{{\rm rad}}_\ell u_\ell(\xi)=
\frac{1}{1-\xi}\left\{\frac{\ell(\ell+d-2)(1-\xi)}{\xi}
-\frac{1}{\xi}({\cal D}^2+d{\cal D})+\frac{2(1-\xi)}{\xi}{\cal D}
+\mu^2\right\}u_\ell(\xi)=\nu^2_\ell \,u_\ell(\xi). 
\label{rad2}
\end{equation}
Using the ansatz
\begin{equation}
u_\ell(\xi)=\xi^{\frac{\ell}{2}}(1-\xi)^{\frac{\Delta_+}{2}}{\cal F}(\xi),  \qquad \Delta_\pm:=\frac{d}{2}\pm\bar{\alpha},\quad\bar{\alpha}:=\sqrt{
\frac{d^2}{4}+\mu^2}\geq0,    
\label{ans}
\end{equation}  
we can verify that ${\cal F}(\xi)$ must satisfiy the hypergeometric
equation with parameters
\begin{equation}
a=\frac{\Delta_++\ell-\nu_\ell}{2},\qquad\qquad
b=\frac{\Delta_++\ell+\nu_\ell}{2},\qquad\qquad c=\ell+\frac{d}{2}.
\end{equation}  

\subsection{Boundary conditions}

At this point it is necessary to discuss boundary conditions. First of all,
we notice that 
the point
$y^1=y^2=\dots=y^d=0$ is just as any other point in AdS (and can be transformed
to any other point) therefore $\psi_{\ell\underline{m}}$ must be analytic at
$y^i=0$. Since $y^\ell Y_{\ell\underline{m}}(\Omega)$ is a polynomial in
$y^i$ (of order $\ell$), we have to require 
$u_\ell(y)=y^\ell f(y^2)$ near $y=0$ with an analytic $f(y^2)$. 
Since the hypergeometric equation with
$c=\ell+\frac{d}{2}$ has two linearly independent solutions, one is
constant at $\xi=0$ (this is given by the hypergeometric function), the other
is singular like ${\cal F}(\xi)\sim\left({1}/{\xi}\right)^{\ell-1+d/2}$,
we conclude that the radial solution must be of the form
\begin{equation}
u_\ell(y)={\cal M}_\ell\left(\frac{R^2}{y^2+R^2}\right)^{\frac{\Delta_+}{2}}  
\left(\frac{y^2}{y^2+R^2}\right)^{\frac{\ell}{2}} {}_2F_1\left(a,b;,c;
\frac{y^2}{y^2+R^2}\right),
\label{hyp1}
\end{equation}  
where ${\cal M}_\ell$ is a normalization constant to be determined later.

Next we  discuss the $y\to\infty$ behaviour of the solutions. We assume
that it is of the form $u_\ell(y)\sim y^{-L}[1+{\rm O}(y^{-2})]$.
In principle the domain of definition may consist of several such classes of
functions with different asymptotic behaviour $L=L_1$, $L_2$, $\cdots$.
Since the solutions are normalizable with respect to the scalar
product (\ref{scalar}),  we require $2L+2 > d$.
Another condition is that
the radial operator defined by (\ref{rad2}) is self-adjoint such that
\begin{equation}
\int_0^\infty{\rm d}y\,y^{d-1}\frac{R^2}{y^2+R^2}u^{(1)}_\ell(y)K^{{\rm rad}}_\ell  
u^{(2)}_\ell(y)= 
\int_0^\infty{\rm d}y\,y^{d-1}\frac{R^2}{y^2+R^2}u^{(2)}_\ell(y)K^{{\rm rad}}_\ell  
u^{(1)}_\ell(y),  
\end{equation}  
where we have to ensure that the boundary
terms (emerging from an integration by part) do not contribute. 
For this condition, we find that
\begin{itemize}

\item
If $u_\ell^{(1)}$ and $u_\ell^{(2)}$ belong to the same class then the
self-adjointness conditions require   $2L_1+2 > d$ and $2L_2+2 >d$,
which is the same as coming from normalizability.

\item
If $u_\ell^{(1)}$ and $u_\ell^{(2)}$ belong to different classes then the
self-adjointness condition becomes   $L_1+L_2 > d$. 

\end{itemize}

\subsection{Spectrum and eigenfunctions}

Let us assume (temporarily) that $\bar{\alpha}$ is not integer. Then using
identities satisfied by the hypergeometric function we can write our solution
(\ref{hyp1}) in an alternative form
\begin{equation}
\begin{split}
&u_\ell(y)={\cal M}_\ell\Gamma\left(\ell+{d}/{2}\right)
\left(\frac{y^2}{y^2+R^2}\right)^{\frac{\ell}{2}}\times\\
&\Bigg\{\left(\frac{R^2}{y^2+R^2}\right)^{\frac{\Delta_+}{2}}
\frac{\Gamma(-\bar{\alpha})}{\Gamma\left(\frac{\ell+\Delta_--\nu_\ell}{2}\right) 
\Gamma\left(\frac{\ell+\Delta_-+\nu_\ell}{2}\right)}{}_2F_1\left(
\frac{\ell+\Delta_+-\nu_\ell}{2},\frac{\ell+\Delta_++\nu_\ell}{2};1+\bar{\alpha};
\frac{R^2}{y^2+R^2}\right)\\
&+\left(\frac{R^2}{y^2+R^2}\right)^{\frac{\Delta_-}{2}}
\frac{\Gamma(\bar{\alpha})}{\Gamma\left(\frac{\ell+\Delta_+-\nu_\ell}{2}\right) 
\Gamma\left(\frac{\ell+\Delta_++\nu_\ell}{2}\right)}{}_2F_1\left(
\frac{\ell+\Delta_--\nu_\ell}{2},\frac{\ell+\Delta_-+\nu_\ell}{2};1-\bar{\alpha};
\frac{R^2}{y^2+R^2}\right)\Bigg\}.
\label{hyp2}
\end{split}    
\end{equation}  
The first term has asymptotic exponent $L_1=\Delta_+$ and the second terms has
$L_2=\Delta_-$. Since $\Delta_+ + \Delta_- =d$, the second condition $L_1+L_2>d$ can never be satisfied.
This means that both terms cannot
simultaneously be present in (\ref{hyp2}). Since
$2\Delta_++2=d+2+2\bar{\alpha}>d$,   
the first term is always normalizable. On the other hand,
$2\Delta_-+2=d+2-2\bar{\alpha}>d $ 
is satisfied only if $\bar{\alpha}<1$.

\subsubsection{$\Delta_+$ case}

The second term in  (\ref{hyp2}) is absent if we choose 
\begin{equation}
\frac{\Delta_++\ell-\nu_\ell}{2}=-n\quad n=0,1,\cdots,\qquad\qquad
\nu_\ell=\nu_{n\ell}=\Delta_++\ell+2n , 
\end{equation}  
since the inverse Gamma function in front of the second term then
vanishes. In this case the first term simplifies to
\begin{equation}
u_{n\ell}(y)={\cal M}_{n\ell}(-1)^n\frac{P_n(\bar{\alpha}+1)}{P_n(\ell+d/2)}
\left(\frac{y^2}{y^2+R^2}\right)^{\frac{\ell}{2}}
\left(\frac{R^2}{y^2+R^2}\right)^{\frac{\Delta_+}{2}}
{}_2F_1\left(-n,\Delta_++\ell+n;1+\bar{\alpha};\frac{R^2}{y^2+R^2}\right).
\label{eq:sol1}
\end{equation} 
We see that the limit $\bar{\alpha}\to{{\rm integer}}$ is smooth.

\subsubsection{$\Delta_-$ case}

If we choose
\begin{equation}
\frac{\Delta_-+\ell-\nu_\ell}{2}=-n\quad n=0,1,\cdots,\qquad\qquad
\nu_\ell=\nu_{n\ell}=\Delta_-+\ell+2n,  
\end{equation}  
the first term in (\ref{hyp2}) vanishes and the second term
becomes
\begin{equation}
u_{n\ell}(y)={\cal M}_{n\ell}(-1)^n\frac{P_n(1-\bar{\alpha})}{P_n(\ell+d/2)}
\left(\frac{y^2}{y^2+R^2}\right)^{\frac{\ell}{2}}
\left(\frac{R^2}{y^2+R^2}\right)^{\frac{\Delta_-}{2}}
{}_2F_1\big(-n,\Delta_-+\ell+n;1-\bar{\alpha};\frac{R^2}{y^2+R^2}\big).
\label{eq:sol2}
\end{equation}

\subsubsection{Final form of the solution}

The possible range of asymptotic exponents (which later become conformal
weights) is
$\frac{d-2}{2} < \Delta$.  
If we introduce the parameters
\begin{equation}
\alpha=\Delta-\frac{d}{2}\quad (\alpha>-1;\quad \bar{\alpha}=\vert\alpha\vert),
\qquad\quad \beta=\ell+\frac{d}{2}-1,  
\end{equation}  
the solutions \eqref{eq:sol1} and \eqref{eq:sol2}  
can be uniformly written as
\begin{equation}
u_{n\ell}(y)={\cal N}_{n\ell}\,\xi^{\ell/2}(1-\xi)^{\Delta/2}P_n^{(\alpha,\beta)}(x),  
\qquad\qquad \nu_{n\ell}=\Delta+\ell+2n.
\end{equation}  
where
\begin{equation}
{\cal N}_{n\ell}=(-1)^n\frac{n!}{P_n(\ell+d/2)}{\cal M}_{n\ell},\qquad\qquad
x=2\xi-1,  
\end{equation}  
and $P_n^{(\alpha,\beta)}(x)$ is the Jacobi polynomial given by
\begin{equation}
P_n^{(\alpha,\beta)}(x)=\frac{\Gamma(n+\alpha+1)}{n! \Gamma(\alpha+1)}
{}_2F_1\left(-n,\alpha+\beta+1+n;\alpha+1;\frac{1-x}{2}\right).
\end{equation}
Using the known
orthogonality properties of the Jacobi polynomials, we can make
our set of solutions orthonormal:
\begin{equation}
\int_0^\infty{\rm d}y\,y^{d-1}\frac{R^2}{y^2+R^2}u_{n\ell}(y)\,u_{m\ell}(y)=
\delta_{nm}.  
\end{equation}  
This requirement fixes the normalization constants as
\begin{equation}
{\cal N}_{n\ell}^2=\frac{2\nu_{n\ell}}{R^d}\,\frac{n!\Gamma(n+\alpha+\beta+1)}
{\Gamma(n+\alpha+1)\Gamma(n+\beta+1)}.
\label{normN}
\end{equation}  
For later use we note that
\begin{equation}
y\to\infty:\qquad\quad u_{n\ell}(y)\approx\frac{P_n(\alpha+1)}{n!}{\cal N}_{n\ell}
\left(\frac{R}{y}\right)^\Delta,  
\end{equation}  
\begin{equation}
y\to0:\qquad\quad u_{n\ell}(y)\approx(-1)^n\frac{P_n(\beta+1)}{n!}{\cal N}_{n\ell}
\left(\frac{y}{R}\right)^\ell.  
\end{equation}  

To summarize, we have found the expansion of the free scalar on the AdS
background in terms of mode functions
\begin{equation}
f_{n\ell\underline{m}}(t,y,\Omega)={\rm e}^{-i\nu_{n\ell}t}u_{n\ell}(y)
Y_{\ell\underline{m}}(\Omega)  
\end{equation}  
for all possible boundary conditions/conformal weights.

\section{Derivation of the bulk reconstruction for $\Delta>d-1$ with odd $d$}
\label{appC}

We evaluate the integral \eqref{Dw1}, using
the hypergeometric function identity (valid for odd $d$) 
\begin{equation}
{}_2F_1(1,d/2;1+\alpha;1/z)=\frac{2\alpha z}{2-d}\,{}_2F_1(1,1-\alpha;2-d/2;z)
+\frac{\Gamma(1-d/2)\Gamma(1+\alpha)}{\Gamma(\Delta-d+1)}
\left(-\frac{1}{z}\right)^{-d/2}(1-z)^{\Delta-d},
\label{hyp}
\end{equation}  
where the first term is regular except for a cut starting at $z=1$. Around the
branch point $z=1$, its behaviour is
\begin{equation}
{\rm regular}\ +\ {\rm const.}\,(1-z)^{\Delta-d}.
\end{equation}  
When calculating the integral of this first term  in \eqref{Dw1}, we can shrink our contour so
that it becomes a very small circle around the branch point $z=1$, and then,
its contribution vanishes in our case (\ref{case}), because a value of the
integral gets smaller and smaller as our integral contour  gets smaller and smaller.

The second term has a cut starting already at $z=0$. The contour can be shrunken
so that it becomes just the unit circle, since the singularity around the
second branch point $z=1$ is an integrable one for (\ref{case}). 
%
After a change of integration variable $z=-{\rm e}^{-2iu}$,
the integral along the unit circle becomes
\begin{equation}
D(w)=\frac{1}{\pi\Omega_d}\frac{\Gamma(1-d/2)\Gamma(1+\alpha)}
{\Gamma(\Delta-d+1)}
\int_{-\pi/2}^{\pi/2}{\rm d}uB\left(-w{\rm e}^{-2iu}\right)
{\rm e}^{-i\Delta u}(2\cos u)^{\Delta-d}.
\end{equation}  
Thus we obtain
\begin{equation}
{\rm e}^{-i\Delta t}D\left({\rm e}^{-2it}\right)=\xi
\int_{t-\pi/2}^{t+\pi/2}{\rm d}u\,{\rm e}^{-i\Delta u} B\left(-{\rm e}^{-2iu}\right)
[2\cos(t-u)]^{\Delta-d}
\end{equation}  
with overall constant $\xi$ in \eqref{eq:xi}.
If we repeat the whole calculation for $D_1$,  we have
\begin{equation}
{\rm e}^{i\Delta t}D_1\left({\rm e}^{2it}\right)=\xi
\int_{t-\pi/2}^{t+\pi/2}{\rm d}u\,{\rm e}^{i\Delta u} B_1\left(-{\rm e}^{2iu}\right)
[2\cos(t-u)]^{\Delta-d}.
\end{equation}
We can simply add the two contributions to arrive at
(\ref{calAC}).

\section{Geometry of the AdS space}
\label{appB}

\subsection{Geodesics}

An important feature of the geometry of AdS space is that $Y$ and $x$ can be connected with
a past directed light-like geodesic if
\begin{equation}
\tilde t=T_1,\qquad\quad T_1=t-\frac{\pi}{2}+\Psi,\qquad\quad \Psi=\arcsin
[(\tanh\rho)\,  n\cdot {\tilde n}].
\label{T1}
\end{equation}
Similarly, $Y$ and $x$ can be connected with
a future directed light-like geodesic if
\begin{equation}
\tilde t=T_2,\qquad\quad T_2=t+\frac{\pi}{2}-\Psi.
\label{T2}
\end{equation}
Finally, $Y$ and $x$ can be connected with
a space-like geodesic if $T_1<\tilde t<T_2$.

\subsection{Infinitesimal transformations}

There is a symmetry action by the isometry grop SO$(d,2)$ on AdS space. The
transformed point will be denoted by $gY$, where $g$ is the group element and
$Y$ is transformed to $gY$. Since it is a group action, the relation
$g_2(g_1Y)=(g_2g_1)Y$ is satisfied.
The corresponding group action in the Hilbert space is given by the unitary
operators $U(g)$ satisfying $U(g_2)U(g_1)=U(g_2g_1)$, under which 
a scalar field $\Phi(Y)$ transforms as $U^\dagger(g)\Phi(Y)U(g)=\Phi(g^{-1}Y)$.
The symmetry group acts also on boundary points by conformal transformations as
$x\to gx$ satisfying $g_2(g_1x)=(g_2g_1)x$.
A primary scalar field ${\cal O}(x)$ transforms under the conformal transformation by the unitary operator as
\begin{equation}
U^\dagger(g){\cal O}(x)U(g)=J^\Delta(g^{-1},x){\cal O}(g^{-1}x), 
\quad J(g,x):=\left\vert\det\,\frac{\partial(gx)^A}{\partial x^B} \right\vert^{\frac{1}{d}} 
= \frac{1}{J(g^{-1},gx)}.
\label{Jid}
\end{equation}

The infinitesimal version of the symmetry transformations are given by
\begin{equation}
\delta \Phi=-\delta Y^\mu\partial_\mu\Phi,\quad\quad
\delta {\cal O}=-\delta x^A\partial_A{\cal O}-\frac{\omega\Delta}{d}
{\cal O}, \quad  \omega:=\frac{\partial\delta x^A}{\partial x^A}.
\end{equation}
The infinitesimal parameters of the SO$(d,2)$ transformations are $E^{AB}=
-E^{BA}$, $A,B=0,D,i$ ($i=1,\dots,d$). The infinitesimal bulk transformations
are given explicitly by
\beqa
\delta t&=&-E^{0D}+\tanh\rho(
 n \cdot E^0\sin t +  n \cdot  E^D\cos t),
\qquad  n \cdot  E^0:=n^i E^{i0},
\quad n \cdot E^D:=n^i E^{iD},\nn \\
\delta \rho&=&- n \cdot E^0\cos t + n \cdot E^D\sin t,\nn \\
\delta n^i&=&E^{ij}n^j+\coth\rho(n \cdot E^0\,n^i-E^{i0})\cos t
-\coth\rho(n \cdot E^D\,n^i-E^{iD})\sin t.
\label{delta}
\eeqa
The boundary (conformal) version of the above is
\begin{equation}
\begin{split}
\delta \tilde t&=-E^{0D}+
 {\tilde n} \cdot E^0\sin \tilde t
+  {\tilde n} \cdot E^D\cos \tilde t,\\
\delta \tilde n^i&=E^{ij}\tilde n^j+( {\tilde n}
\cdot E^0\,\tilde n^i-E^{i0})\cos \tilde t
-( {\tilde n} \cdot E^D\,\tilde n^i-E^{iD})\sin \tilde t.
\end{split}
\label{bardelta}
\end{equation}
For the boundary points there is no $\tilde\rho$ coordinate, but for later use 
we keep the notation $\delta \tilde\rho=- {\tilde n} \cdot E^0\cos \tilde t
+ {\tilde n} \cdot E^D\sin \tilde t$.
Also for later use we note that
$\delta J=- \delta\tilde\rho={\omega}/{d}$.

\section{Bulk reconstruction for $\Delta>d-1$ at generic bulk points}
\label{appE}

In this appendix we construct the building blocks $I(Y,x)$ and $T(Y,x)$, which
are necessary to complete the bulk reconstruction for generic bulk points
discussed in section \ref{sec2}.

\subsection{The explicit form of $I(Y,x)$}

The infinitesimal version of the first requirement in (\ref{Ifun}) is
$\delta I=(\delta J)I=\frac{\omega}{d}\,I=-\delta \tilde\rho \,I$.
We start from the invariant function depending on two bulk points in AdS given by
\begin{equation}
S=\cosh\rho \cosh\tilde\rho\cos(t-\tilde t)-\sinh\rho\sinh\tilde\rho\,
 n\cdot{\tilde n},\qquad\quad \delta S=0.
\end{equation}
For large $\tilde\rho$, the second point goes to the boundary and we have
approximately
\begin{equation}
S\approx\frac{1}{4}{\rm e}^{\tilde\rho}\,I,\qquad\quad I=2[\cosh\rho
\cos(t-\tilde t)-\sinh\rho \, n\cdot{\tilde n}].
\end{equation}
In this limit, the infinitesimal variation gives $0=\delta S=\frac{1}{4}{\rm e}^{\tilde\rho}(\delta{\tilde\rho}\,I+\delta I)$, which leads to
$\delta I=-\delta{\tilde\rho}\,I$.
The first requirement in its infinitesimal form is thus satisfied by this $I$, which also satisfies
 the second requirement since
$I(Y_o,x)=2\cos\tilde t$ for $t=\rho=0$.

\subsection{The explicit form of $T(Y,x)$}

A natural guess is to take $T(Y,x)=\xi\Theta(X_1)\Theta(X_2)$,
where $X_1=\tilde t-T_1$ and $X_2=T_2-\tilde t$.
The second requirement in (\ref {Tfun}) is satisfied with this choice since
$T(Y_o,x)=\xi\Theta\left(\tilde t+\frac{\pi}{2}\right)\Theta\left(\frac{\pi}{2}-\tilde t\right)$.

The infinitesimal variation of $X_1$ can be calculated using the formulas given
in (\ref{delta}) and (\ref{bardelta}). After some calculation, we obtain
\beqa
\delta X_1&=&2\sin\frac{X_1}{2}\Big\{
 n\cdot E^0\Big(-\frac{\tanh\rho}{\cos\Psi}\sin\frac{\tilde t+T_1}{2}\Big)
+ n\cdot E^D\Big(
-\frac{\tanh\rho}{\cos\Psi}\cos\frac{\tilde t+T_1}{2}\Big)\nn \\
&+&  {\tilde n}\cdot E^0\Big(\tan\Psi\sin\frac{\tilde t+T_1}{2}
+\cos\frac{\tilde t+T_1}{2}\Big)
+  {\tilde n}\cdot E^D\Big(\tan\Psi\cos\frac{\tilde t+T_1}{2}
-\sin\frac{\tilde t+T_1}{2}\Big)\Big\},\nn \\
\eeqa
so that $X_1=0$ implies $\delta X_1=0$.
Thus $\Theta(X_1)$ is invariant: $\Theta(X_1) = \Theta(X_1+\delta X_1)$
for infinitesimal changes.
We also observe that $\displaystyle\lim_{X_1\to0}{\delta X_1}/{X_1}=\delta\lambda$, 
where
\begin{equation}
\begin{split}
\delta\lambda&= n\cdot E^0
\Big(-\frac{\tanh\rho}{\cos\Psi}\sin T_1\Big)
+n\cdot E^D
\Big(-\frac{\tanh\rho}{\cos\Psi}\cos T_1\Big)\\
&+ {\tilde n}\cdot E^0
(\tan\Psi \sin\tilde t+\cos\tilde t)
+{\tilde n}\cdot E^D
(\tan\Psi \cos\tilde t-\sin\tilde t).
\end{split}
\end{equation}

We can make similar calculations and draw similar conclusions for $X_2$.
For its infinitesimal variation, we obtain
\beqa
\delta X_2&=&-2\sin\frac{X_2}{2}\Big\{
n\cdot E^0\Big(
-\frac{\tanh\rho}{\cos\Psi}\sin\frac{\tilde t+T_2}{2}\Big)
+ n\cdot E^D\Big(
-\frac{\tanh\rho}{\cos\Psi}\cos\frac{\tilde t+T_2}{2}\Big)\nn \\
&+& {\tilde n}\cdot E^0\Big(\tan\Psi\sin\frac{\tilde t+T_2}{2}
-\cos\frac{\tilde t+T_2}{2}\Big)
+  {\tilde n}\cdot E^D\Big(\tan\Psi\cos\frac{\tilde t+T_2}{2}
+\sin\frac{\tilde t+T_2}{2}\Big)\Big\}, \nn \\
\eeqa
so that $X_2=0$ implies $\delta X_2=0$.
Finally $\displaystyle\lim_{X_2\to0}{\delta X_2}/{X_2}=\delta\bar\lambda$, 
where
\begin{equation}
\begin{split}
\delta\bar\lambda&= n\cdot E^0
\Big(\frac{\tanh\rho}{\cos\Psi}\sin T_2\Big)
+ n\cdot E^D
\Big(\frac{\tanh\rho}{\cos\Psi}\cos T_2\Big)\\
&- {\tilde n}\cdot E^0
(\tan\Psi \sin\tilde t-\cos\tilde t)
- {\tilde n}\cdot E^D
(\tan\Psi \cos\tilde t+\sin\tilde t).
\end{split}
\end{equation}

\section{Bulk reconstruction for $\Delta>d-2$}
\label{appD}

We separate the bulk and boundary fields, ${\cal A}(t)$ and ${\cal C}(t)$, into
positive/negative frequency parts, ${\cal A}_+(t)/{\cal A}_-(t)$ and
${\cal C}_+(t)/{\cal C}_-(t)$, which are given by the two terms of
(\ref{Apm}) and (\ref{Cpm}), respectively. 
Using these definitions, we have the identity
\begin{equation}
{\rm e}^{-i\Delta t}B({\rm e}^{-2it})={\rm e}^{-i\frac{\Delta\pi}{2}}
{\cal C}_+(t-\pi/2)={\rm e}^{i\frac{\Delta\pi}{2}}{\cal C}_+(t+\pi/2).
\end{equation}
Thus (\ref{Dw}) leads to
\beqa
{\cal A}_+(t)&=&\frac{1}{\Omega_d}{\rm e}^{-\frac{i\Delta\pi}{2}}{\cal C}_+(t-\pi/2)
+\xi\int_{-\pi/2}^0{\rm d}u [2\cos(u)]^{\Delta-d}\{{\cal C}_+(u+t)
-{\rm e}^{-i(u+\pi/2)\Delta}{\cal C}_+(t-\pi/2)\}\nn \\
&+&\xi\int_0^{\pi/2}{\rm d}u [2\cos(u)]^{\Delta-d}\{{\cal C}_+(u+t)
-{\rm e}^{-i(u-\pi/2)\Delta}{\cal C}_+(t+\pi/2)\}.
\eeqa
Next by adding and subtracting an integral proportional to $k_+$ for the first
integral and $k_-$ for the second integral, where
\begin{equation}
k_\pm=\xi\int_0^{\pi/2}{\rm d}u(2\cos u)^{\Delta-d}
[1-{\rm e}^{\pm i\Delta(u-\pi/2)}],
\end{equation}
which are convergent for $\Delta>d-2$, we obtain
\begin{equation}
\begin{split}
{\cal A}_+(t)&=\left\{\frac{1}{\Omega_d}+{\rm e}^{\frac{i\Delta\pi}{2}} k_+
+{\rm e}^{\frac{-i\Delta\pi}{2}} k_-\right\}
{\rm e}^{-\frac{i\Delta\pi}{2}}{\cal C}_+(t-\pi/2)\\
&+\xi\int_{t-\pi/2}^t{\rm d}u [2\cos(u-t)]^{\Delta-d}\{{\cal C}_+(u)
-{\cal C}_+(t-\pi/2)\}\\
&+\xi\int_t^{t+\pi/2}{\rm d}u [2\cos(u-t)]^{\Delta-d}\{{\cal C}_+(u)
-{\cal C}_+(t+\pi/2)\}.
\end{split}
\end{equation}
Analogously, repeating the calculation with $D_1(w)$, $B_1(w)$, ${\cal A}_-$
and ${\cal C}_-$, we have
\begin{equation}
{\rm e}^{i\Delta t}B_1({\rm e}^{2it})={\rm e}^{i\frac{\Delta\pi}{2}}
{\cal C}_-(t-\pi/2)={\rm e}^{-i\frac{\Delta\pi}{2}}{\cal C}_-(t+\pi/2)
\end{equation}
and
\begin{equation}
\begin{split}
{\cal A}_-(t)&=\left\{\frac{1}{\Omega_d}+{\rm e}^{\frac{i\Delta\pi}{2}} k_+
+{\rm e}^{\frac{-i\Delta\pi}{2}} k_-\right\}
{\rm e}^{\frac{i\Delta\pi}{2}}{\cal C}_-(t-\pi/2)\\
&+\xi\int_{t-\pi/2}^t{\rm d}u [2\cos(u-t)]^{\Delta-d}\{{\cal C}_-(u)
-{\cal C}_-(t-\pi/2)\}\\
&+\xi\int_t^{t+\pi/2}{\rm d}u [2\cos(u-t)]^{\Delta-d}\{{\cal C}_-(u)
-{\cal C}_-(t+\pi/2)\}.
\end{split}
\end{equation}

These results can be further simplified by using the following two identities.
\begin{equation}
\begin{split}
{\cal C}(t&-\pi/2)+{\cal C}(t+\pi/2)
={\cal C}_+(t-\pi/2)+{\cal C}_-(t-\pi/2)+{\cal C}_+(t+\pi/2)+
{\cal C}_-(t+\pi/2)\\
&=2\cos\frac{\Delta\pi}{2}\big\{{\rm e}^{-\frac{i\Delta\pi}{2}}{\cal C}_+(t-\pi/2)+
{\rm e}^{\frac{i\Delta\pi}{2}}{\cal C}_-(t-\pi/2)\big\},
\end{split}
\end{equation}
\begin{equation}
\begin{split}
\int_0^{\pi/2}{\rm d}u&(2\cos u)^A\left\{\cos\frac{B\pi}{2}-\cos Bu\right\}
=\frac{\pi}{2}\,\Gamma(1+A)\Big\{\frac{\cos\frac{B\pi}{2}}{\Gamma^2(1+A/2)}
-\frac{1}{\Gamma(1+\frac{A-B}{2})\Gamma(1+\frac{A+B}{2})}\Big\},
\end{split}
\end{equation}
for $A>-2$, $B=A+d$, $d=3,5,7,\cdots$.
Using the second identity we find
\begin{equation}
\begin{split}
{\rm e}^{\frac{i\Delta\pi}{2}}k_+&+{\rm e}^{-\frac{i\Delta\pi}{2}}k_-=
\xi\int_0^{\pi/2}{\rm d}u(2\cos u)^{\Delta-d}\big\{2\cos\frac{\Delta\pi}{2}-
2\cos\Delta u\big\}
=\frac{1}{\Omega_d}\big\{\eta\cos\frac{\Delta\pi}{2}-1\big\}.
\end{split}
\end{equation}
Finally, adding ${\cal A}_+$ and ${\cal A}_-$ we obtain the final result
(\ref{calAC2}), valid
for the extended range $\Delta>d-2$.

\section{Details of the derivation of bulk reconstruction for small integer
$\Delta$}  
\label{appF}

\subsection{Bulk reconstruction for $\Delta=d-1$ at  generic bulk points}

We will look for
a solution to the infinitesimal form of the first requirement, $\delta D=-\delta J\,D=\delta\tilde\rho\,D$.
For the ansatz $D(Y,x)=\xi_o[f(Y,x)\delta_o(X_1)+g(Y,x)\delta_o(X_2)]$,
the second requirement is satisfied if $f(Y_o,x)=g(Y_o,x)=1$.
Using the delta function relations
\begin{equation}
\begin{split}
\delta_o(X_1+\delta X_1)&=\delta_o(X_1+\delta\lambda X_1)=(1-\delta\lambda)
\delta_o(X_1),\\
\delta_o(X_2+\delta X_2)&=\delta_o(X_2+\delta\bar\lambda X_2)
=(1-\delta\bar\lambda)
\delta_o(X_2)
\end{split}
\end{equation}
we see that the first requirement is equivalent to
\begin{equation}
\frac{\delta f}{f}=\delta\lambda+\delta\tilde\rho\qquad\quad
({\rm on\ the}\ X_1=0\ {\rm hyperplane}), 
\label{delf}
\end{equation}
\begin{equation}
\frac{\delta g}{g}=\delta\bar\lambda+\delta\tilde\rho\qquad\quad
({\rm on\ the}\ X_2=0\ {\rm hyperplane}). 
\label{delg}
\end{equation}
Considering 
\begin{equation}
\sin T_1=\sin\Psi\sin t-\cos\Psi\cos t,\qquad\quad
\cos T_1=\cos\Psi\sin t+\sin\Psi\cos t,
\end{equation}
we have
\begin{equation}
\begin{split}
\delta\lambda&=-\tanh\rho\,n\cdot E^0(\tan\Psi\sin t-\cos t)
-\tanh\rho\, n\cdot E^D(\sin t+\tan\Psi\cos t)\\
&+{\tilde n}\cdot E^0(\tan\Psi\sin\tilde t+\cos\tilde t)
+{\tilde n}\cdot E^D(\tan\Psi\cos\tilde t-\sin\tilde t)\\
&=-\tan\Psi(\delta t+E^{0D})-\tanh\rho\,
\delta\rho+\tan\Psi(\delta {\tilde t}
+E^{0D})-\delta\tilde\rho\\
&=\tan\Psi\delta(\tilde t-t)-\tanh\rho\,\delta\rho-\delta\tilde\rho.
\end{split}
\end{equation}
Under the condition $X_1=0$
\begin{equation}
\delta\lambda+\delta\tilde\rho=-\tanh\rho\,\delta\rho+\tan\Psi\delta\Psi=
-\frac{\delta\cosh\rho}{\cosh\rho}-\frac{\delta\cos\Psi}{\cos\Psi}.
\end{equation}
Thus the solution to (\ref{delf}) is obtained by $f=\dfrac{1}{{\cal R}}$,
where ${\cal R}=\cosh\rho\cos\Phi$ is given in \eqref{eq:defR}.

Similarly we have
\begin{equation}
\delta\bar\lambda
=\tan\Psi\delta(t-\tilde t)-\tanh\rho\,\delta\rho-\delta\tilde\rho,
\end{equation}
which, under the condition $X_2=0$, implies
\begin{equation}
\delta\bar\lambda+\delta\tilde\rho
=-\tanh\rho\,\delta\rho+\tan\Psi\delta\Psi=
-\frac{\delta{\cal R}}{{\cal R}}.
\end{equation}
We find that (\ref{delg}) has the same solution, $g=\dfrac{1}{{\cal R}}$.
Thus the result for the complete kernel function is given by (\ref{kernelD}).

\subsection{Bulk reconstruction for $\Delta=d-2$ at  generic bulk points}

Let us first concentrate on the $X_1$ part of $D_2$.
Motivated by the $\Delta=d-1$ result we take the ansatz
\begin{equation}
D_2(Y,x)=-\tilde\xi_o[f_2(Y,x)\delta^\prime_o(X_1)+p_2(Y,x)\delta_o(X_1)]
\ +\ X_2\ {\rm part}.  
\end{equation}
The first requirement is $\delta D_2 = -2\delta J D_2= 2\delta\bar\rho D_2$.
The second requirement will be satisfied if $f_2(Y_o,x)=1$ and $p_2(Y_o,x)=0$.

\subsubsection{Delta function identities}

We start from the well-known delta function relation
\begin{equation}
\delta_o(f(x))=\frac{1}{\vert f^\prime(0)\vert}\delta_o(x),
\end{equation}
where we assume that the only zero of $f(x)$ is at $x=0$. Then
\begin{equation}
\begin{split}  
\int{\rm d}&x\delta^\prime_o(f(x)){\cal F}(x)=\int{\rm d}x\delta^\prime_o(f(x))
f^\prime(x)\,\frac{{\cal F}(x)}{f^\prime(x)}
=-\int{\rm d}x\delta_o(f(x))\Big[\frac{{\cal F}(x)}{f^\prime(x)}\Big]^\prime\\
&=-\frac{{\cal F}^\prime(0)}{\vert f^\prime(0)\vert f^\prime(0)}
+\frac{{\cal F}(0)f^{\prime\prime}(0)}{\vert f^\prime(0)\vert {f^\prime}^2(0)}.
\end{split}
\end{equation}
This gives the delta function identity
\begin{equation}
\delta^\prime_o(f(x))=\frac{1}{\vert f^\prime(0)\vert f^\prime(0)}
\delta^\prime_o(x)+\frac{f^{\prime\prime}(0)}{\vert f^\prime(0)\vert {f^\prime}^2(0)}
\delta_o(x).
\end{equation}
We will apply the above identities to the infinitesimal variation of $X_1$:
\begin{equation}
\delta X_1=\varepsilon_0X_1+\varepsilon_1 X^2_1+{\rm O}(X_1^3).
\label{vareps}
\end{equation}
In this case, $\delta^\prime_o(X_1+\delta X_1)=(1-2\varepsilon_0)\delta^\prime_o(X_1)+
2\varepsilon_1\delta_o(X_1)$.  
The infinitesimal change of the delta function and its derivative is thus 
\begin{equation}
\delta[\delta_o(X_1)]=-\varepsilon_0\delta_0(X_1),\qquad  
\delta[\delta^\prime_o(X_1)]=-2\varepsilon_0\delta^\prime_0(X_1)
+2\varepsilon_1\delta_o(X_1).
\end{equation}
Later we will also use the identity $X_1\delta^\prime_o(X_1)=-\delta_o(X_1)$. 

\subsubsection{Expansions}

For later use we now calculate and simplify the expansion coefficients in
(\ref{vareps}) and in the expansion of $\delta\tilde\rho$ (defined under
(C.6)): $\delta\tilde\rho=r_0+r_1X_1+{\rm O}(X^2_1)$.
Using
\begin{equation}
\tilde t=T_1+X_1,\qquad \frac{\tilde t+T_1}{2}=T_1+\frac{X_1}{2}
\end{equation}
we find from (D.3)
\begin{equation}
\begin{split}
\varepsilon_0&=n\cdot E^0\big(-\frac{\tanh\rho}{\cos\Psi}\sin T_1\big)
+n\cdot E^D\big(-\frac{\tanh\rho}{\cos\Psi}\cos T_1\big)\\
&+\tilde n\cdot E^0(\tan\Psi\sin T_1+\cos T_1)
+\tilde n\cdot E^D(\tan\Psi\cos T_1-\sin T_1),
\end{split}    
\end{equation}
\begin{equation}
\begin{split}
2\varepsilon_1&=n\cdot E^0\big(-\frac{\tanh\rho}{\cos\Psi}\cos T_1\big)
+n\cdot E^D\big(\frac{\tanh\rho}{\cos\Psi}\sin T_1\big)\\
&+\tilde n\cdot E^0(\tan\Psi\cos T_1-\sin T_1)
-\tilde n\cdot E^D(\tan\Psi\sin T_1+\cos T_1).
\end{split}    
\end{equation}
For the expansion of $\delta\tilde\rho$ we find
\begin{equation}
r_0=-\tilde n\cdot E^0\cos T_1+\tilde n\cdot E^D\sin T_1,\quad
r_1=\tilde n\cdot E^0\sin T_1+\tilde n\cdot E^D\cos T_1.        
\end{equation}
Using the relations (F.4), (C.5), (C.6) we make the following
calculations (for later use).
\begin{equation}
\begin{split}
\varepsilon_0+r_0&=\tanh\rho\, n\cdot E^0(\cos t-\tan\Psi\sin t)
-\tanh\rho\, n\cdot E^D(\sin t+\tan\Psi\cos t)\\
&+\tan\Psi[\tilde n\cdot E^0\sin(\tilde t-X_1)
+\tilde n\cdot E^D\cos(\tilde t-X_1)]\\
&=-\tan\Psi(\delta t+E^{0D})-\tanh\rho\delta\rho+\tan\Psi[(\delta\tilde t+E^{0D})
+r_0 X_1]+{\rm O}(X^2_1)\\  
&=-\tanh\rho\delta\rho+\tan\Psi(\delta\Psi+\delta X_1+r_0 X_1)+{\rm O}(X^2_1)\\
&=\tan\Psi\delta\Psi-\tanh\rho\delta\rho+\tan\Psi(\varepsilon_0+r_0)X_1+
{\rm O}(X^2_1).
\end{split}    
\end{equation}
\begin{equation}
\begin{split}
2\varepsilon_1+2r_1&=-\tanh\rho\,n\cdot E^0(\sin t+\tan\Psi\cos t)
-\tanh\rho\,n\cdot E^D(\cos t-\tan\Psi\sin t)\\
&+\tan\Psi(\tilde n\cdot E^0\cos\tilde t-\tilde n\cdot E^D\sin\tilde t)
+(\tilde n\cdot E^0\sin\tilde t+\tilde n\cdot E^D\cos\tilde t)
+{\rm O}(X_1)\\
&=-(\delta t+E^{0D})+\tan\Psi\tanh\rho\delta\rho-r_0\tan\Psi+(\delta\tilde t
+E^{0D})+{\rm O}(X_1)\\
&=\delta\Psi+\tan\Psi\tanh\rho\delta\rho-r_0\tan\Psi+{\rm O}(X_1).
\end{split}    
\end{equation}
After temporarily dropping the $-\tilde\xi_o$ factor the first requirement
(for the $X_1$ part) reads
\begin{equation}
\delta[f_2\delta^\prime_o(X_1)+p_2\delta_o(X_1)]=2\delta\tilde\rho
[f_2\delta^\prime_o(X_1)+p_2\delta_o(X_1)],   
\end{equation}
which can be expanded as
\begin{equation}
\delta f_2\delta^\prime_o +
f_2[-2\varepsilon_0\delta^\prime_o+2\varepsilon_1
\delta_o]+\delta p_2\delta_o-\varepsilon_0 p_2\delta_o
=2r_0 f_2\delta^\prime_o-2r_1 f_2\delta_o+2r_0 p_2\delta_o.
\end{equation}
Reducing to zero and dividing by $f_2$ we get
\begin{equation}
\left[\frac{\delta f_2}{f_2}-2(\varepsilon_0+r_0)\right]\delta^\prime_o
+\delta_o\left[2(\varepsilon_1+r_1)+\frac{\delta p_2}{f_2}-(2r_0+\varepsilon_0)
\frac{p_2}{f_2}\right]=0.  
\end{equation}
Thus the condition that the leading term multiplying $\delta^\prime_o$ vanishes leads to $f_2={\cal R}^{-2}$.
Introducing the parametrization $p_2=f_2\omega_2$,
the requirement that the coefficient of $\delta_o$ vanishes becomes
\begin{equation}
2\tan\Psi(\varepsilon_0+r_0)+2(\varepsilon_1+r_1)+2(\varepsilon_0+r_0)\omega_2
+\delta\omega_2-(2r_0+\varepsilon_0)\omega_2=0.
\end{equation}
This can be simplified to
\begin{equation}
\begin{split}
&2\tan\Psi(\varepsilon_0+r_0)
+\delta\Psi+\tan\Psi\tanh\rho\delta\rho-r_0 \tan\Psi
+\delta\omega_2+\varepsilon_0\omega_2\\  
&=\varepsilon_0\tan\Psi+\tan^2\Psi\delta\Psi+\delta\Psi+\delta\omega_2+
\varepsilon_0\omega_2=\frac{\delta\Psi}{\cos^2\Psi}+\delta\omega_2+
\varepsilon_0(\omega_2+\tan\Psi)=0.
\end{split}    
\end{equation}
Now it is easy to see that $\omega_2=-\tan\Psi$ solves this equation, and indeed $ p_2(Y_o,x)=0$.

The calculation of the $X_2$ part is completely analogous.

\section{BDHM relation for $\Delta=d-1$}
\label{appG}

Although the BDHM relation \eqref{eq:BDHM} \cite{Banks:1998dd} is one of the staring points of our derivation for \eqref{eq:generic_d-1},
it is instructive to check it directly from the final formula   \eqref{eq:generic_d-1}.

By writing 
\beqa
\epsilon:={1\over \sinh\rho}, \ \tanh\rho={1\over\sqrt{1+\epsilon^2}}, \
n_y\cdot n_x := \cos\gamma, \ {\cal R}={\sqrt{\epsilon^2+\sin^2\gamma}\over \epsilon},
\eeqa
we should show 
\beqa
O(t,\Omega_y) &=& \lim_{\epsilon\to0} {\xi_o \over \epsilon^{\Delta-1}}\int d\Omega\, {1\over  \sqrt{\epsilon^2+\sin^2\gamma}}
\left[O(T_1,\Omega) +O(T_2,\Omega)\right],
\eeqa
where
\beqa
T_1=t-{\pi\over 2} +\Psi, \ T_1=t+{\pi\over 2} -\Psi, \  \Psi:= \sin^{-1} \left({\cos\gamma\over \sqrt{1+\epsilon^2}}\right).
\eeqa
Since the BDHM relation \eqref{eq:BDHM} is a linear mapping, it is enough to verify it mode by mode.
For the operator $A_{nl\underline{m}}$, 
what we need to show is
\beqa
Y_{l \underline{m}}(\Omega_y) =
 \lim_{\epsilon\to0} {\xi_o \over \epsilon^{\Delta-1}}\int d\Omega\, { Y_{l \underline{m}}(\Omega) \over\sqrt{\epsilon^2+\sin^2\gamma}}\left[ e^{i\nu_{nl} (\pi/2-\Psi)}+e^{-i\nu_{nl} (\pi/2-\Psi)}\right].
 \label{eq:show1}
\eeqa

Without loss of generality, we can take $\Omega_y\sim n_0:=(1,0,0,\cdots, 0)$, so that $\gamma=\theta$ and $\Psi={\pi\over 2} -\theta$.
From the property of the hyper-spherical harmonics 
\beqa
Y_{l m \underline{\hat m}}(\Omega) = N^d_{lm\underline{\hat m}} (\sin\theta)^{m}\, C^{\alpha + m}_{l-m}(\cos\theta) Y_{m \underline{\hat m}}(\hat \Omega) ,
\quad \alpha:={d\over 2}-1, \ \underline{m}:= m\underline{\hat m},
\eeqa  
where $\hat \Omega$ is a solid angle of the $d-2$ dimensional sphere,
we obtain
\beqa
Y_{l \underline{m}}(\Omega_y\sim n_0) &=& \delta_{\underline{m},\underline{0}} Y_{l\underline{0}}(n_0), \quad
Y_{l\underline{0}}(n_0) = N^d_{l\underline{0}}  C^{\alpha}_{l}(1) a_{d-1}
\eeqa
since $Y_{\underline{0}}(\hat\Omega) = a_{d-1} =1/\sqrt{ \Omega_{d-1}}$.
Furthermore, using $d\Omega = \sin^{d-2}\theta \ d\theta\, d\hat\Omega$,
we have
\beqa
\xi_o\int d\Omega\,  Y_{l\underline{m}}(\Omega) F(\theta)&=&\delta_{{\underline{m}},\underline{0}}N^d_{l\underline{0}} \xi_o a_{d-1}\Omega_{d-1} \int_0^\pi d\theta  \sin^{d-2}\theta F(\theta)
 C^\alpha_l(\cos\theta) ,
\eeqa
where
\beqa
F(\theta) :=  {1\over \epsilon^{\Delta-1}} { 1 \over\sqrt{\epsilon^2+\sin^2\theta}}
 \left[ e^{i\nu_{nl} \theta}+e^{-i\nu_{nl} \theta}\right].
\eeqa
We then evaluate the integral given by
\beqa
X:=\lim_{\epsilon\to 0}  {1\over \epsilon^{\Delta-1}} \int_0^\pi d\theta  { \sin^{d-2}\theta  \over\sqrt{\epsilon^2+\sin^2\theta}}
 C^\alpha_l(\cos\theta) 
 \left[ e^{i\nu_{nl} \theta}+e^{-i\nu_{nl} \theta}\right].
\eeqa
For odd $d=2s+3$ with $\Delta=d-1$, we have 
\beqa
X &=& \lim_{\epsilon\to 0}  {1\over \epsilon^{d-2}} \int_{-\pi}^\pi d\theta  { \sin^{d-2}\theta  \over\sqrt{\epsilon^2+\sin^2\theta}}
 C^\alpha_l(\cos\theta) e^{i\nu_{nl} \theta}.
\eeqa
By rewriting
\beqa
 {1\over \epsilon^{d-2}} { \sin^{d-2}\theta  \over\sqrt{\epsilon^2+\sin^2\theta}} &=& D_\epsilon (\theta) + \delta D_\epsilon(\theta),
\eeqa
where
\beqa
D_\epsilon(\theta)& :=&  {1\over \epsilon^{d-2}} { \sin^{d-2}\theta  \over\sqrt{\epsilon^2+\sin^2\theta}} -\delta D_\epsilon(\theta) , \\
\delta D_\epsilon(\theta) &:=&{1\over \epsilon} \sum_{k=0}^s {(2k-1)!! \over (2k) !! }(-1)^k \left({\sin^2\theta\over \epsilon^2}\right)^{s-k},
\eeqa
we see that
\beqa
\int_{-\pi}^\pi d\theta\, \delta D_\epsilon(\theta) C^\alpha_l(\cos\theta) e^{i\nu_{nl} \theta} =0, 
\eeqa
since $\delta D_\epsilon(\theta) $ and $C^\alpha_l(\cos\theta)$ are polynomials of $e^{i\theta}$, $e^{-i\theta}$ of order $(d-3)$ and $l$, respectively,
while $\nu_{nl} =d-1+2n+l$.

Since $D_\epsilon(\theta)$ satisfies
\beqa
\lim_{\epsilon\to 0} D_\epsilon(\theta) =\left\{
\begin{array}{llc}
 O(\epsilon)  &\to 0,   &\mbox {for $\sin\theta\not =0$},   \\
   (-1)^{s+1} O(\epsilon^{-1})  & \to (-1)^{s+1}\infty,   & \mbox {for $\sin\theta=0$},  \\
\end{array}
\right. 
\eeqa
and $D_\epsilon(\pi-\theta)=D_\epsilon(\theta)$, we  conclude
\beqa
\lim_{\epsilon\to 0} D_\epsilon(\theta) &=& A(s) \left[\delta(\theta) +\delta(\pi-\theta)\right].
\eeqa

\subsection{Calculation of $A$}
We write
\beqa
A &=&\lim_{\epsilon\to 0}( A_0 + A_1), \quad
A_0 := \int_0^\pi d\theta\,  {1\over \epsilon^{d-2}} { \sin^{d-2}\theta  \over\sqrt{\epsilon^2+\sin^2\theta}},\
A_1:= -\int_0^\pi  d\theta\,\delta D_\epsilon(\theta). 
\eeqa
Thus, $A_1$ becomes
\beqa
A_1&=& -{\sqrt{\pi}\over\epsilon^{2s+1}} \sum_{k=0}^s \left[ \epsilon^{2k}(-1)^k {(2k-1)!!\over (2k)!!}{\Gamma(s-k+{1\over 2})\over \Gamma(s-k+1)}\right].
\eeqa
The calculation of $A_0$ is more involved. Making a change of variables as  $\cos\theta =\sqrt{1+\epsilon^2}\sin w$, we obtain
\beqa
A_0&=& \int_{-a}^a dw\, \left[\cos^2 w -\epsilon^2\sin^2 w\right]^s, 
\eeqa
where
\beqa
a&:=& \sin^{-1}{1\over \sqrt{1+\epsilon^2}}=\cos^{-1}{\epsilon\over \sqrt{1+\epsilon^2}}
={\pi\over 2} -{\epsilon \over 1+\epsilon^2}\sum_{n=0}^\infty {(2n)!!\over (2n+1)!!}\left({\epsilon^2\over 1+\epsilon^2}\right)^n\nn \\
&=& {\pi\over 2} -\epsilon\left[ 1 -{\epsilon^2\over 3} +{3\epsilon^4\over15} +O(\epsilon^6)\right].
\eeqa
We calculate $A_0$ for $s=0,1,2$ as follows.
\beqa
\epsilon A_0(s=0) &=& 2a ,\\
\epsilon^3 A_0(s=1) &=& a(1-\epsilon^2) + \sin (2a){(1+\epsilon^2)\over 2}
=a(1-\epsilon^2)+\epsilon,\\
\epsilon^5 A_0(s=2) &=&a{(3 -2\epsilon^2+ 3\epsilon^4)\over 4}+\sin(2a){(1-\epsilon^4)\over 2}
+\sin(4a){(1+2\epsilon^2+\epsilon^4)\over 16}\nn \\
&=& {1\over 4}\left[ a (3 -2\epsilon^2+ 3\epsilon^4) + 3\epsilon(1-\epsilon^2)\right],
 \eeqa
 where we use 
 \beqa
 \sin(2a) ={2\epsilon\over 1+\epsilon^2}, \quad
 \sin(4a) ={4\epsilon(\epsilon^2-1)\over (1+\epsilon^2)^2}.
 \eeqa
On the other hands,
\beqa
A_1(s=0)&=& -{\pi\over\epsilon}, \\
A_1(s=1) &=& -{\pi\over 2\epsilon^3}(1-\epsilon^2),\\
A_1(s=2) &=& -{\pi \over 8\epsilon^5}\left(3 -2\epsilon^2+3\epsilon^4\right).
\eeqa
By combining these, we obtain
\begin{equation}
A(s=0) = -2, \qquad
A(s=1) ={4\over 3}, \qquad
A(s=2) =-{16\over 15}.
\end{equation}

\subsection{The result}
The right-hand side of \eqref{eq:show1} now becomes
\begin{equation}
A(s)N_{l\underline{0}}^d \xi_o a_{d-1} \Omega_{d-1}\left[C_l^\alpha(1) +(-1)^l C_l^\alpha(-1)\right]\delta_{\underline{m},\underline{0}}=
2 A(s)  \xi_o \Omega_{d-1} Y_{l\underline{m}}(n_0), 
\end{equation}
where we have used $C^\alpha_l(-1)=(-1)^l C^\alpha_l(1)$.
Since
\begin{equation}
2\xi_o \Omega_{d-1} = {(-1)^{s+1} (2s+1) !!\over 2^{s+1} s!} = -{1\over 2}, {3\over 4}, -{15\over 16}, 
\end{equation}
for $s=0,1,2$, which implies $A(s) 2\xi_o \Omega_{d-1}=1$,
so that the BDHM relation  \eqref{eq:show1} holds for $s=0,1,2$.
We expect in general 
\beqa
A(s) = (-1)^{s+1} {2^{s+1} s!\over (2s+1)!!},
\eeqa
for all non-negative integer $s$, which we verified up to $s=10$ by Mathematica.



\bibliographystyle{JHEP}
\bibliography{Flow}

\end{document}